\documentclass[lettersize,journal]{IEEEtran}
\usepackage{amsmath,amsfonts}
\usepackage{algorithm}
\usepackage{array}
\usepackage[caption=false,font=normalsize,labelfont=sf,textfont=sf]{subfig}
\usepackage{textcomp}
\usepackage{stfloats}
\usepackage{url}
\usepackage{verbatim}
\usepackage{graphicx}
\usepackage{amssymb}
\usepackage{algorithm}      
\usepackage{algorithmic}      
\usepackage{eqparbox}
\usepackage{xargs}   
\usepackage{spconfpdkul}
\usepackage{pgfplots}   
\usepackage{ifthen}
\usepackage[acronym, shortcuts]{glossaries}  

\usepackage[
  backend=biber,
  style=ieee,
  sorting=none,  
  doi=false,
  isbn=false,
  url=false,
  eprint=false
]{biblatex}

\newacronym{wasn}{WASN}{wireless acoustic sensor network}
\newacronym{danse}{DANSE}{distributed adaptive node-specific signal estimation}
\newacronym{dansep}{TI-DANSE$^+$}{topology-independent distributed adaptive node-specific signal estimation PLUS[TO BE CHANGED]}
\newacronym{gevddansep}{TI-GEVD-DANSE$^+$}{topology-independent GEVD-based distributed adaptive node-specific signal estimation PLUS[TO BE CHANGED]}
\newacronym{gevd_or_dansep}{TI-(GEVD-)DANSE$^+$}{topology-independent (GEVD-based) distributed adaptive node-specific signal estimation PLUS[TO BE CHANGED]}
\newacronym{tdanse}{T-DANSE}{tree-DANSE}
\newacronym{tidanse}{TI-DANSE}{topology-independent distributed adaptive node-specific signal estimation}
\newacronym{tigevddanse}{TI-GEVD-DANSE}{topology-independent GEVD-based distributed adaptive node-specific signal estimation}
\newacronym{ti_or_gevddanse}{TI-(GEVD-)DANSE}{topology-independent (GEVD-based) distributed adaptive node-specific signal estimation}
\newacronym{gevddanse}{GEVD-DANSE}{GEVD-based distributed adaptive node-specific signal estimation}
\newacronym{gevd}{GEVD}{generalized eigenvalue decomposition}
\newacronym{lmmse}{LMMSE}{linear minimum mean square error}
\newacronym{mmse}{MMSE}{minimum mean square error}
\newacronym{mwf}{MWF}{multichannel Wiener filter}
\newacronym{gevdmwf}{GEVD-MWF}{GEVD-based multichannel Wiener filter}
\newacronym{scm}{SCM}{spatial covariance matrix}
\newacronym{snr}{SNR}{signal-to-noise ratio}
\newacronym{mse}{MSE}{mean square error}
\newacronym{vad}{VAD}{voice activity detector}
\newacronym{mst}{MST}{minimum spanning tree}
\newacronym{gsbcd}{GSBCD}{Gauss-Seidel block-coordinate descent}
\newacronym{ao}{AO}{alternating optimization}
\newacronym{mmut}{MMUT}{multiple max-$|\Uk|$ trees}
\newacronym{mc}{MC}{Monte-Carlo}
\newacronym{ac}{AC}{algebraic connectivity}
\newacronym{fc}{FC}{fully connected}
\newacronym{stft}{STFT}{short-term Fourier transform}
\newacronym{dft}{DFT}{discrete Fourier transform}
\newacronym{stoi}{STOI}{short-term objective intelligibility}
\newacronym{dof}{DoF}{degree of freedom}
\newacronym{ssn}{SSN}{speech-shaped noise}
\newacronym{sro}{SRO}{sampling rate offset}
\newacronym{sto}{STO}{sampling time offset}
\newacronym{pesq}{PESQ}{Perceptual Evaluation of Speech Quality}
\newacronym{rir}{RIR}{room impulse response}

\DeclareUnicodeCharacter{2212}{−}
\usepgfplotslibrary{groupplots,dateplot}
\usetikzlibrary{patterns,shapes.arrows,external}
\pgfplotsset{compat=newest}

\begin{document}

\title{Fast-Converging Distributed Signal Estimation in\\ Topology-Unconstrained Wireless Acoustic Sensor Networks}

\author{PAUL DIDIER\IEEEauthorrefmark{1}, TOON VAN WATERSCHOOT\IEEEauthorrefmark{1}, SIMON DOCLO\IEEEauthorrefmark{2,3}, J{\"O}RG BITZER\IEEEauthorrefmark{3}, AND MARC MOONEN\IEEEauthorrefmark{1}\\
\IEEEmembership{
  \IEEEauthorrefmark{1}STADIUS Division, Department of Electrical Engineering (ESAT), KU Leuven, 3001 Leuven, Belgium\\
  \IEEEauthorrefmark{2}Signal Processing Group, Department of Medical Physics and Acoustics and Cluster of Excellence Hearing4all,\\Carl von Ossietzky Universit{\"a}t Oldenburg, Oldenburg, Germany\\
  \IEEEauthorrefmark{3}Fraunhofer IDMT, Project Group Hearing, Speech and Audio Technology, Oldenburg, Germany
}
\thanks{This research work was carried out in the frame of the Research Council KU Leuven: C14-21-0075 "A holistic approach to the design of integrated and distributed digital signal processing algorithms for audio and speech communication devices" and the European Union's Horizon 2020 research and innovation programme under the Marie Skłodowska-Curie Grant Agreement No. 956369: ``Service-Oriented Ubiquitous Network-Driven Sound — SOUNDS''. This paper reflects only the authors' views and the Union is not liable for any use that may be made of the contained information. The scientific responsibility is assumed by the authors.}}

\markboth{Submitted to IEEE Transactions on Signal and Information Processing over Networks}%
{Didier \MakeLowercase{\textit{et al.}}: Fast-Converging Distributed Signal Estimation in Topology-Unconstrained WASNs}


\maketitle

\begin{abstract}
  This paper focuses on distributed node-specific signal estimation in topology-unconstrained wireless acoustic sensor networks (WASNs) where sensor nodes only transmit fused versions of their local sensor signals.
  For this task, the topology-independent (TI) distributed adaptive node-specific signal estimation (DANSE) algorithm (TI-DANSE) has previously been proposed. It converges towards the centralized signal estimation solution in non-fully connected and time-varying network topologies.
  However, the applicability of TI-DANSE in real-world scenarios is limited due to its slow convergence. The latter results from the fact that, in TI-DANSE, nodes only have access to the in-network sum of all fused signals in the WASN.
  We address this low convergence speed issue by introducing an improved TI-DANSE algorithm, referred to as TI-DANSE$^+$.
  \change{
  The TI-DANSE$^+$ algorithm outperforms TI-DANSE in terms of convergence speed by letting the updating node use each partial in-network sum of fused signals (coming from its neighbors) separately, when updating its estimation parameters. In this way, the number of available degrees of freedom in the optimization problem at the updating node is increased, leading to faster convergence.}

  \change{
  This separate use of incoming partial in-network sums} is further exploited by combining TI-DANSE$^+$ with a tree-pruning strategy that maximizes the number of neighbors at the updating node.
  In fully connected WASNs, it is observed that TI-DANSE$^+$ converges as fast as the original DANSE algorithm (the latter only defined for fully connected WASNs) while using peer-to-peer data transmission instead of broadcasting and thus saving communication bandwidth. If link failures occur, the convergence of TI-DANSE$^+$ towards the centralized solution is preserved without any change in its formulation.
  Altogether, the proposed TI-DANSE$^+$ algorithm can be viewed as an all-round alternative to DANSE and TI-DANSE which (i) merges the advantages of both, (ii) reconciliates their differences into a single formulation, and (iii) shows advantages of its own in terms of communication bandwidth usage. The convergence properties \change{
  and signal estimation performance} of TI-DANSE$^+$ are demonstrated through speech enhancement experiments in simulated topology-unconstrained WASNs.
\end{abstract}



\section{Introduction}\label{sec:intro}

The ever-increasing ubiquity of devices equipped of acoustic sensors (microphones) in our environment enables the creation of \glspl*{wasn} composed of multiple wirelessly interconnected devices (also called multi-sensor ``nodes''), e.g., smartphones, laptops, or hearing aids~\cite{bertrand_applications_2011, alias_review_2019}. Such \glspl*{wasn} can be used as distributed systems, as opposed to centralized systems in which processing is exclusively performed at a fusion center where all audio signals are aggregated. Each node in a \gls*{wasn} can receive, locally process, and transmit signals to other nodes. By wirelessly exchanging signals, devices can access information about sound sources of interest even when these are far from them.
The real-world deployment of distributed systems remains a challenging task as several practical requirements must be fulfilled such as, e.g., low communication bandwidth usage to avoid network congestion, robustness to dynamic network topologies as nodes may drop in and out of the \gls*{wasn}, and quick adaptivity~\cite{plata-chaves_heterogeneous_2017}. These challenges are best addressed early on, i.e., at the application layer in the design of distributed algorithms, which is the focus of this paper.

\subsection{State-of-the-art and Related Works}\label{subsec:related_works}

Research works in the field of distributed algorithms have so far largely focused on parameter estimation, i.e., estimating a single, static or slowly varying parameter vector (e.g., the dimensions of a room, the location of sound sources, or features of a speech signal) of interest to all nodes in the \gls*{wasn}~\cite{jin-jun_xiao_distributed_2006,schizas_distributed_2007,sayed_diffusion_2013}. 
\change{
This has traditionally be done using, e.g., consensus-based techniques~\cite{xiaoFastLinearIterations2004,mateosDistributedSparseLinear2010,doostmohammadianConsensusBasedDistributedEstimation2022,doostmohammadianDistributedTargetTracking2025}.} 
However, in many applications, nodes need to estimate entire signals, i.e., samples of an underlying incoming nonstationary process such as a speech signal~\cite{ying_jia_distributed_2006, doclo_reduced-bandwidth_2009}, which may also be node-specific.
\change{
Although consensus-based strategies may be used in cases where all nodes aim to estimate the same signal, node-specific signal estimation problems can be efficiently tackled via spatial filtering-based solutions that exploit the correlation between the signals recorded at different nodes. In spatial filtering-based methods, the target signal of any given node is estimated by performing a linear combination of all signals available at that node, where the weights of that combination are found as solution to an optimization problem involving the signals recorded by all nodes in the network.}

To achieve the same performance as a centralized system \change{
through distributed spatial filtering}, a naive solution may be to allow all nodes to broadcast all of their sensor signals \change{
to all other nodes in the \gls*{wasn}}. This would, however, result in intolerably high communication bandwidth usage, especially when many nodes are involved. \change{
This motivated the development of the \gls*{danse} algorithm~\cite{bertrand_distributed_2010}, in which nodes first combine their local sensor signals into low-dimensional (i.e., fused) signals before transmitting them, thereby reducing bandwidth usage. It was shown that the \gls*{danse} algorithm converges towards the centralized solution when the target signals of all nodes in the \gls*{wasn} emerge from a common set of latent sound sources.}
Using the \gls*{danse} algorithm in an \gls*{fc} \gls*{wasn}, the same signal estimation performance as an equivalent centralized system can be obtained iteratively, without relying on a fusion center.

The \gls*{danse} algorithm is designed to function solely in \gls*{fc} \glspl*{wasn}, i.e., networks where all nodes can directly transmit data to all other nodes. This motivated the development of alternative formulations, first for \glspl*{wasn} with static tree-topologies~\cite{bertrand_distributed_2011_TDANSE} and static mixed-topologies~\cite{szurley_distributed_2015}, then for any, possibly time-varying topologies via the topology-independent (TI) \gls*{danse} algorithm (\glsunset{tidanse}\gls*{tidanse})~\cite{szurley_topology-independent_2017}.
This algorithm achieves convergence in any topology-unconstrained \gls*{wasn} by pruning the network to a tree topology at each algorithmic iteration and building the in-network sum of all fused signals, which is then used for node-specific signal estimation.

Even though \gls*{tidanse} enables distributed signal estimation in any \gls*{wasn}, it converges slowly towards the optimum~\cite{szurley_topology-independent_2017,didier2024tigevddanse}. This slow convergence is an inherent characteristic of \gls*{tidanse} since, at each iteration, the updating node only has access to the global in-network sum of all fused signals to update its filters.
The number of degrees of freedom (\glspl*{dof}) in the local \gls*{tidanse} optimization problem is indeed much smaller than for \gls*{danse} in an equivalent \gls*{fc} \gls*{wasn}, where nodes have access to all individual fused signals from all other nodes. Letting $K$ represent the number of nodes in the \gls*{wasn} and $Q$ the number of fused signals computed at and exchanged between each node, an updating node in \gls*{danse} has access to $Q(K-1)$ signals transmitted by other nodes while an updating node in \gls*{tidanse} has access to only $Q$ signals transmitted by other nodes. The resulting issue is particularly noticeable when deploying \gls*{tidanse} in an \gls*{fc}, or nearly \gls*{fc} \gls*{wasn}, where convergence to the centralized solution is then significantly slower than that of the \gls*{danse} algorithm. This drawback of \gls*{tidanse} severely hinders its applicability in real-world scenarios that require fast adaptivity.

\subsection{Contributions}\label{subsec:contributions}

In this paper, a distributed algorithm termed \glsunset{dansep}\gls*{dansep} is proposed to resolve the slow convergence issue of \gls*{tidanse}. The \gls*{dansep} algorithm allows all nodes in any unconstrained and possibly time-varying \gls*{wasn} topology to reach the same signal estimation performance as if they were part of a centralized system, while only exchanging low-dimensional fused versions of their local sensor signals.
The \gls*{dansep} algorithm achieves this without compromising on convergence speed when the network is \gls*{fc}, i.e., reaching the same convergence speed as \gls*{danse} in \gls*{fc} \glspl*{wasn}.

Letting $B\geq 1$ denote the number of direct neighbors of the updating node, the key enabling principle of \gls*{dansep} lies in the fact that the updating node uses $B$ partial in-network sums of fused signals coming from each of its neighbors \textit{separately} to update its estimation parameters. \change{
This is a major difference with respect to \gls*{tidanse}, where the partial sums are summed into a global sum, which is then used to perform signal estimation.}
\change{An updating node in \gls*{dansep} thus has access to $QB$ fused signals, against $Q$ for an updating node in \gls*{tidanse}.
Since $B=K{-}1$ in a \gls*{fc} \gls*{wasn}, \gls*{dansep} converges as fast as \gls*{danse} while requiring less communication bandwidth at every iteration since \gls*{dansep} relies on peer-to-peer data transmission from all nodes to the updating node and back, while \gls*{danse} relies on broadcasting from every node to every other node and back.}

The use of partial in-network sums of fused signals coming from different branches in a tree-topology \gls*{wasn} has been explored in~\cite{musluoglu_distributed_2023} for a general signal fusion problem. However, the solution proposed there requires each non-updating node to solve an additional optimization problem \change{
which, to be solved, requires the estimation at each iteration of one additional set of signal statistics per non-updating node. This greatly limits the practical applicability of~\cite{musluoglu_distributed_2023}.}

Since the number of \glspl*{dof} in the local optimization problem at an updating node in \gls*{dansep} depends on the number $B$ of direct neighbors of the updating node, the choice of \gls*{wasn} tree-pruning strategy is crucial and impacts the convergence speed of the algorithm.
In this paper, a tree-pruning strategy that maximizes the number of \glspl*{dof} at the updating node is proposed.
Using this strategy, simulated experiments show that \gls*{dansep} converges as fast as \gls*{danse} in \gls*{fc} \glspl*{wasn} and faster than \gls*{tidanse} in any non-\gls*{fc} \gls*{wasn} topology. In general, it is observed that the more connections exist between nodes of the original non-pruned \gls*{wasn}, the faster \gls*{dansep} can converge.

The \gls*{dansep} algorithm can be viewed as an all-round alternative to \gls*{danse} and \gls*{tidanse} which (i) merges the advantages of both, (ii) reconciliates their differences into a single formulation, and (iii) shows advantages of its own in terms of communication bandwidth in \gls*{fc} \glspl*{wasn}.

In~\cite{didierImprovedTopologyIndependentDistributeda}, a brief high-level description of the \gls*{dansep} algorithm was introduced. In this paper, we provide more details, including a communication complexity analysis, \change{
a computational complexity analysis, the addressing of several implementation aspects, and more extensive simulation results including dynamic topologies, \gls*{gevd}-based filter updates, and simulations in a realistic speech enhancement setting where signal statistics are estimated from the available microphone recordings}. Importantly, a convergence proof is provided, formally demonstrating that \gls*{dansep} converges towards the centralized solution in any topology-unconstrained \gls*{wasn}.

\subsection{Paper Organization}

In~\secref{sec:prob_statement}, the signal model is defined and the centralized \gls*{mwf} solution to the node-specific signal estimation problem is given.
In~\secref{sec:dansep}, the \gls*{dansep} algorithm is formally introduced, including practical considerations, and its convergence towards the centralized solution is proven.
In~\secref{sec:res}, simulation results are provided to quantitatively evaluate the performance of \gls*{dansep} for different tree-pruning strategies and to compare its performance with that of \gls*{danse} and \gls*{tidanse}. Finally, conclusions are formulated in~\secref{sec:ccl}. 

\textit{Notation}: Italic letters denote scalars and lowercase boldface letters denote column vectors. Uppercase boldface letters are used for matrices. The superscripts $\cdot^\T$ and $\cdot^\Her$ denote the transpose and Hermitian transpose, respectively. The symbols $\E[\cdot]$, $\|\cdot\|_2$, and $\|\cdot\|_\mathrm{F}$ denote the expected value operator, the Euclidean norm, and the Frobenius norm, respectively. The operator $\mathrm{diag}\{\mathbf{a}\}$ transforms the vector $\mathbf{a}$ in a diagonal matrix with $\mathbf{a}$ on its main diagonal.
The symbol $|\mathcal{A}|$ denotes the number of elements in the set $\mathcal{A}$. The symbol $\mathbf{0}_{A\times B}$ denotes the $A\times B$ all-zeros matrix and $\I[A]$ the $A\times A$ identity matrix.
The set of complex numbers is denoted by $\mathbb{C}$.

\section{Problem Statement}\label{sec:prob_statement}

Consider a \gls*{wasn} composed of $K$ nodes \change{
with ideal communication links\footnote{\change{
Ideal communication links refers to a setting where packet losses, network-related time-delays, and \glspl*{sro} are absent. Although the impact of non-ideal communication links on the performance of the proposed distributed algorithm remains beyond the scope of this paper, it is noted that time-delay and/or \gls*{sro} estimation and compensation mechanisms may be incorporated in the \gls*{dansep} framework described in~\secref{sec:dansep}. For instance, fixed inter-node time-delays and \glspl*{sro} may be compensated for in topology-unconstrained \glspl*{wasn} using methods akin to, e.g.,~\cite{didier_sampling_2023}.}}} deployed in an environment with $S\geq 1$ localized sound sources of interest. The set of all node indices is denoted $\K\triangleq \{1,\dots,K\}$. Node $k\in\K$ has $M_k$ local sensors (microphones) such that the total number of sensors in the \gls*{wasn} is $M\triangleq \sum_{k\in\K} M_k$. In the following, we consider the signals to be complex-valued to allow for frequency-domain representations, e.g., in the \gls*{stft} domain.
The signal $y_{k,m}[t]$ captured at time-frame (e.g., \gls*{stft} frame) $t$ by the $m$-th sensor of node $k$ is modelled as $y_{k,m}[t] =  s_{k,m}[t] + n_{k,m}[t]$, where $s_{k,m}[t]$ and $n_{k,m}[t]$ represent the desired signal component and the noise component, respectively. The noise component $n_{k,m}[t]$ may contain contributions from localized or diffuse noise sources as well as sensor noise. For the sake of conciseness, the time index $[t]$ will be dropped in the following.
All sensor signals of node $k$ are stacked in the column vector $\mathbf{y}_k \triangleq  [y_{k,1},\dots,y_{k,M_k}]^\T\in\C[M_k]$, resulting in the multichannel signal model at node $k$:

\begin{equation}\label{eq:signal_model}
  \yk =  \sk + \nk,
\end{equation}

\noindent
where $\sk$ and $\nk$ are defined in a similar way as $\yk$.
The desired signal component in~\eqref{eq:signal_model} can be further expanded as:

\begin{equation}\label{eq:sk_to_slat}
  \sk =  \Psibk\mathbf{s}^\mathrm{lat},
\end{equation}

\noindent
where $\Psibk\in\C[M_k][S]$ is the steering matrix for node $k$ and $\mathbf{s}^\mathrm{lat}\in\C[S]$ is the latent signal produced by the desired sources.
Equation~\eqref{eq:sk_to_slat} can be interpreted as the frequency-domain equivalent of time-domain convolutions with impulse responses. Convolutive time-domain mixtures can thus be processed as instantaneous per-frequency mixtures in the (short-term) Fourier transform domain. Note that the frame length must be chosen sufficiently large to fully capture the information of the underlying impulse responses.

As in~\cite{bertrand_distributed_2010}, we consider the node-specific signal estimation problem where each node $k\in\K$ aims at estimating its own $Q$-dimensional target signal $\dk$, where $Q\leq M_k\fa k\in\K$. We assume that $Q$ is the same for all nodes. This target signal is a subset of the desired signal component in~\eqref{eq:signal_model}:

\begin{equation}\label{eq:dk_to_sk}
  \dk \triangleq  \Ekk^\T\sk = \Psibkov \mathbf{s}^\mathrm{lat}\in\C[Q],
\end{equation}

\noindent
where $\Ekk\in\{0,1\}^{M_k\times Q}$ is a selection matrix and $\Psibkov \triangleq \Ekk^\T\Psibk\in\C[Q][S]$.
The target signal model from~\eqref{eq:dk_to_sk} is typical in a variety of practical scenarios, e.g., for a binaural hearing aid system striving to estimate a speech signal impinging on a single reference microphone in the left as well as in the right hearing aid (node) while exploiting all microphone signals from both hearing aids (nodes)~\cite{doclo_reduced-bandwidth_2009,cornelis_theoretical_2010}.

Considering the centralized case, where each node has access to all sensor signals in the \gls*{wasn}, the local sensor signal vectors from~\eqref{eq:signal_model} can be themselves stacked into a centralized signal vector $\mathbf{y}\in\C[M]$ as:

\begin{equation}\label{eq:yCentr}
  \mathbf{y} \triangleq  \begin{bmatrix}
    \yk[1]^\T \dots \yk[K]^\T
  \end{bmatrix}^\T.
\end{equation}

\noindent
This vector can itself be split into desired and noise components, denoted by $\mathbf{s}$ and $\mathbf{n}$, respectively. 
The objective for node $k$ is to compute a node-specific filter $\Wk\in\C[M][Q]$ such that $\Wk^\Her\mathbf{y}$ provides an estimate of $\dk$ considering the \gls*{lmmse} criterion. The optimal filter $\hWk$ is obtained by solving:

\begin{equation}\label{eq:lmmseCentr}
  \hWk \triangleq  \underset{\mathbf{W}\in\C[M][Q]}{\mathrm{arg\,min}}\:
  \E[{
      \left\|
          \dk - \mathbf{W}^\Her\mathbf{y}
      \right\|^2_2
  }].
\end{equation}

\noindent
The solution of~\eqref{eq:lmmseCentr} is the \gls*{mwf}:

\begin{equation}\label{eq:centr_mwf}
    \hWk=\left(\Ryy\right)^{-1}\Rydk,
\end{equation}

\noindent
with the spatial covariance matrices (\glsunset{scm}\glspl*{scm}) $\Ryy\triangleq \mathbb{E}\{\mathbf{yy}^\Her\}\in\C[M][M]$ and $\Rydk\triangleq \mathbb{E}\{\mathbf{yd}_k^\Her\}\in\C[M][Q]$. Estimating $\Ryy$ can be done by averaging over the available observations of $\mathbf{y}$, assuming short-term stationarity of the signals. Even though $\dk$ is obviously unavailable at node $k$, $\Rydk$ can be estimated by first assuming that the desired and noise components in~\eqref{eq:signal_model} are uncorrelated, which is often the case in, e.g., speech enhancement scenarios, giving:

\begin{align}\label{eq:Rydk_RssEk}
  \Rydk
  &= \E[\mathbf{y}\dk^\Her]
  = \E[\mathbf{s}\dk^\Her]\\
  &= \E[\mathbf{ss}^\Her]\Ek
  \triangleq  \Rss\Ek,\label{eq:Rydk_RssEk_final}
\end{align}

\noindent
While $\Rss$ itself cannot be directly observed, it can be indirectly estimated as $\Ryy - \Rnn$, where the noise \gls*{scm} $\Rnn \triangleq  \mathbb{E}\{\mathbf{nn}^\Her\}\in\C[M][M]$ can, e.g., be estimated by making use of the on-off characteristics of the desired speech signal and a \gls*{vad} or by using prior knowledge~\cite{zhao2020model, dov2017multimodal}. In practical applications, the \glspl*{scm} may be estimated using exponential averaging as:

\begin{equation}\label{eq:expavg}
  \begin{split}
    \Ryy[t] &= \beta\Ryy[t-1] + (1-\beta)\mathbf{y}[t]\mathbf{y}^\Her[t]\:\:\text{if VAD$[t]=1$},\\
    \Rnn[t] &= \beta\Rnn[t-1] + (1-\beta)\mathbf{y}[t]\mathbf{y}^\Her[t]\:\:\text{if VAD$[t]=0$},
  \end{split}
\end{equation}

\noindent
where $0\ll\beta<1$ is a forgetting factor and VAD$[t]\in\{0,1\}$ denotes the \gls*{vad} value at time-frame $t$.

\section{The TI-DANSE$^+$ Algorithm}\label{sec:dansep}

In~\secref{sec:prob_statement}, the nodes had access to all sensor signals in the \gls*{wasn}, allowing each node to directly compute the centralized \gls*{mwf} via~\eqref{eq:centr_mwf}. In this section, we consider the general case of a topology-unconstrained \gls*{wasn}.
Mathematical notation and semantics used to interpret \glspl*{wasn} as graphs are introduced in~\secref{subsec:graphs}. In~\secref{subsec:dansep_def}, the \gls*{dansep} algorithm is introduced, keeping a consistent notation with respect to \gls*{danse}~\cite{bertrand_distributed_2010}. Its convergence and optimality are proven in~\secref{subsec:convergence}. The importance of the chosen tree-pruning strategy is discussed in~\secref{subsec:pruning_strategy_discussion}. Finally, several aspects related to the practical implementation of \gls*{dansep} are discussed in~\secref{subsec:implementation_aspects}.

\subsection{Wireless Sensor Networks as Graphs}\label{subsec:graphs}

A \gls*{wasn} can be seen as an undirected\footnote{\change{
For simplicity, undirected graphs are considered in this paper. However, in \glspl*{wasn} with asymmetric communication links (i.e., where signals can be transmitted from node $a$ to node $b$ but not from $b$ to $a$), directed graphs may be a more appropriate model. The \gls*{dansep} algorithm may be extended to directed graphs where a path exists between any two nodes following the direction of the edges (i.e., strongly connected graphs). Although this remains out of the scope of the present paper, this may be achieved by accounting for edge directivity in the \gls*{dansep} fusion and diffusion flows definitions, as introduced in following sections.}} graph where the nodes are the vertices and the connections between the nodes are the edges.
The \gls*{wasn} is considered to be connected if data can be exchanged between any two nodes of the network, either directly (neighbor nodes) or via a series of hops through intermediate nodes.
A \gls*{wasn} is \gls*{fc} if every node in the graph is directly connected to every other node.

A \gls*{wasn} has a tree topology with root node $k$ if it contains no cycles. Any node in a tree may have a single parent (neighbor node towards the root) and one or more children (neighbor node(s) away from the root). Nodes with no children are referred to as leaf nodes, and nodes with both children and a parent are referred to as internal nodes. The root node does not have any parent. In a tree, two directions are defined, namely (i) upstream from root towards leaf nodes and (ii) downstream from leaf nodes towards root.
For any node $q\in\K$, the set of upstream neighbor nodes of $q$ is denoted by $\Uk[q]$. The set of all nodes upstream of $q$ is denoted by $\Ukb[q]$; it thus holds that $\Uk[q]\subseteq\Ukb[q],\:\forall\:q\in\K$.

\begin{figure}[h]
  \centering
  \includegraphics[width=.95\columnwidth,trim={0 0 0 0},clip=false]{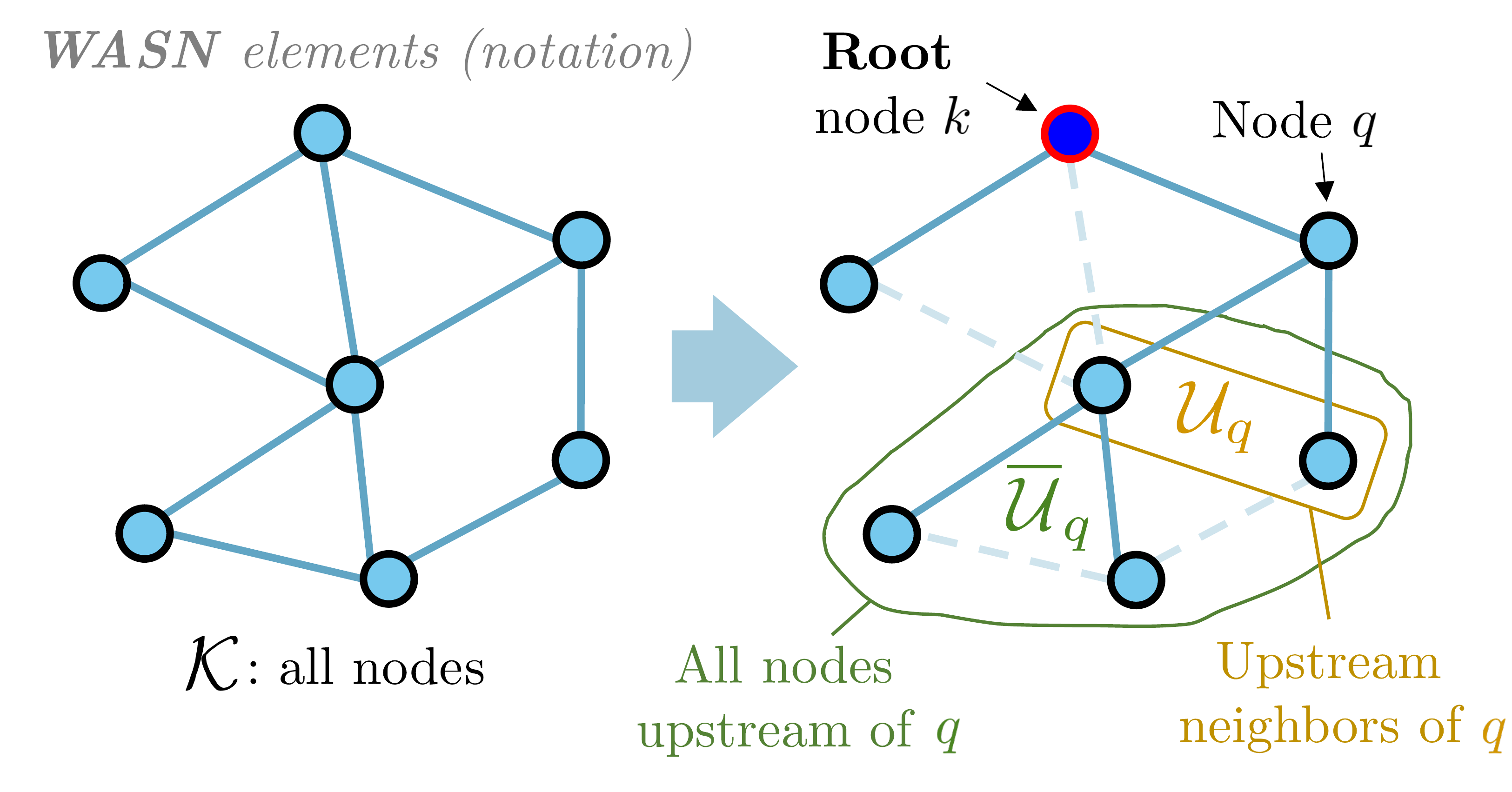}
  \caption{Example pruning of a topology-unconstrained \gls*{wasn} to a tree topology with root $k$, including sets notation for a node $q\in\K\backslash\{k\}$.}
  \label{fig:pruning_example}
\end{figure}

Any connected \gls*{wasn} may be pruned to a tree\footnote{A trivial case being a topology-unconstrained \gls*{wasn} that is already a tree, in which case pruning is irrelevant.} by, e.g., message-passing initiated at the root node~\cite{gallagerDistributedAlgorithmMinimumWeight1983}. An example of arbitrary tree-pruning is given in~\figref{fig:pruning_example}.
Networks with a tree topology are generally easier to manage as there exists only one path between any two nodes. If the objective is to minimize a certain quantity associated to data transmission across the network, e.g., communication bandwidth usage, strategies for finding a \gls*{mst} can be used. Examples of such techniques include Kruskal's algorithm~\cite{kruskal1956shortest} and other methods~\cite{prim1957shortest,chen2009minimax,rab2017improved}. However, \glspl*{mst} represent only a subset of all possible spanning trees for a given \gls*{wasn} and they may not be the ideal choice depending on the targeted application. In~\secref{subsec:pruning_strategy_discussion}, we will define some relevant tree-pruning strategies and experimentally compare the performance of the proposed \gls*{dansep} algorithm when using these strategies in~\secref{subsec:res_static}.

\subsection{Algorithm Definition}\label{subsec:dansep_def}

In this section, we define the \gls*{dansep} algorithm at a given iteration $i$ \change{
for application in a \gls*{wasn} modelled as an undirected connected graph}.
\change{Note that the iteration index $i$ is different from the time index $t$ from~\secref{sec:prob_statement}. In order to estimate the statistics (i.e., the \glspl*{scm}) required within one iteration via time-averaging, data from several time-frames are necessary. This means that index $i$ is typically incremented more slowly than the time-frame index $t$.}
The \change{
  updating node} index $k(i)$ cycles over all node indices in a round-robin fashion.
\footnote{Non-sequential node-updating strategies such as simultaneous or asynchronous node-updating were developed for \gls*{danse} in~\cite{bertrand_distributed_2010-1} but remain outside the scope of this paper.}

At iteration $i$, the \gls*{wasn} is first pruned to a tree with root node $k(i)$, where node $k(i)$ updates its local filters. 
A central aspect of \gls*{tidanse} is retained in \gls*{dansep} in that the pruned tree does not have to be the same at every iteration, consequently making \gls*{dansep} robust to dynamic \gls*{wasn} topologies and link failures.
To account for the iteration-dependent tree, a superscript is added to the sets $\Uk[q]^i$ and $\Ukb[q]^i$ for all $q\in\K$. 
For the sake of notation simplicity, we omit the index $(i)$ in $k(i)$ in the following unless specified, keeping in mind that $k$ denotes the root node at iteration $i$.

Each iteration of the \gls*{dansep} algorithm can be split into three steps: (i) a \textit{fusion flow}, where linear combinations of the local sensor signals (i.e., fused signals) are computed and sent from leaf nodes towards the root node, (ii) the \textit{filter update} at the root node, and (iii) a \textit{diffusion flow}, where some filter coefficients and the target signal estimate at the root node are flooded back through the tree, from the root node towards the leaf nodes.

\subsubsection{Fusion Flow}\label{subsubsec:fusionflow}

\begin{figure}[h]
  \centering
  \includegraphics[width=\columnwidth,trim={0 0 0 0},clip=false]{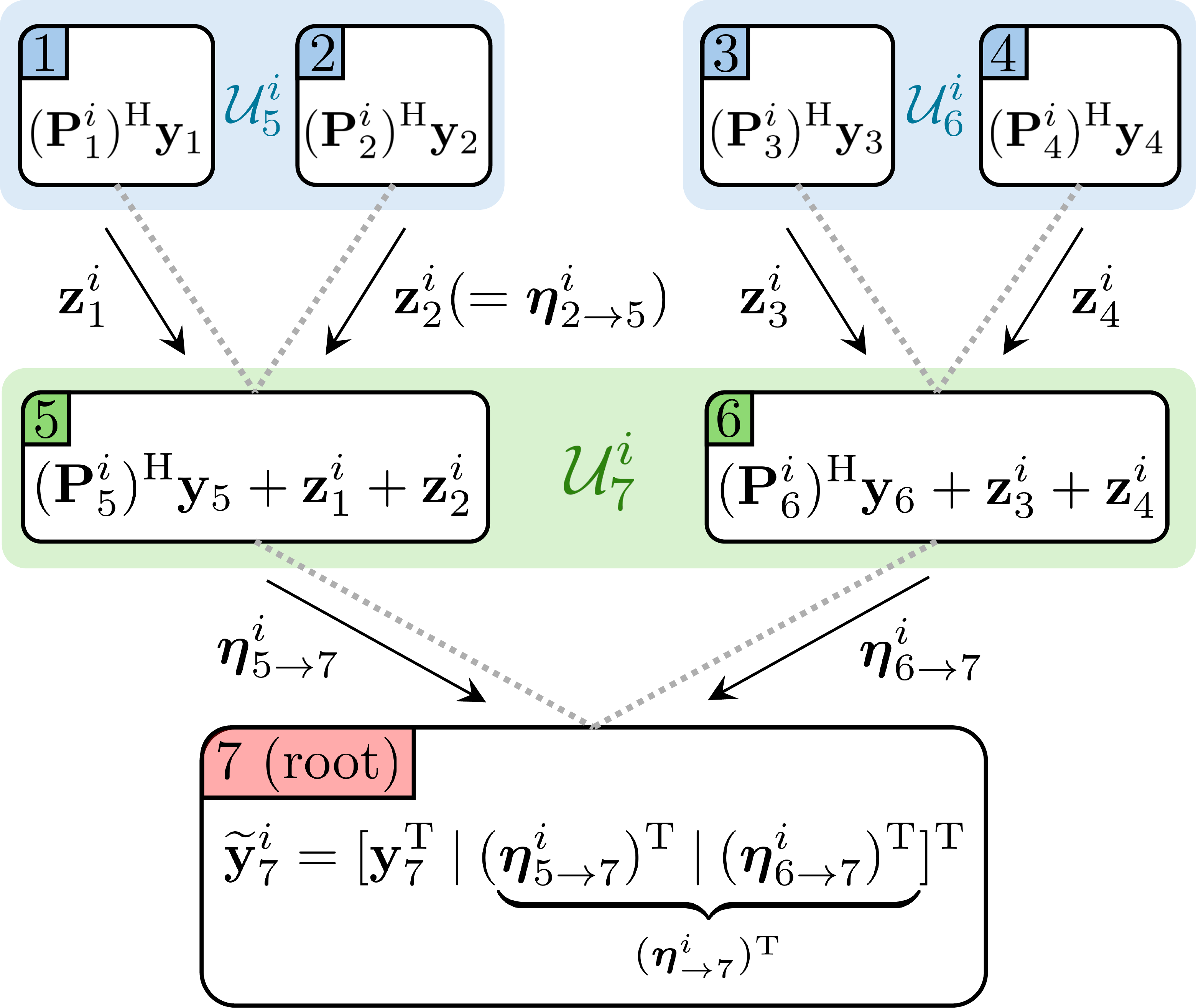}
  \caption{Schematic representation of the \gls*{dansep} fusion flow in a topology-unconstrained \gls*{wasn} with $K=7$ nodes, pruned to a tree with root node index $k=7$. \change{
    Note that, at leaf nodes $l\in\{1,2,3,4\}$, $\zk[l]^i=\ettkq[l][p]^i$ with $p$ the parent of leaf node $l$.
  }}
  \label{fig:fusionflow}
\end{figure}

After tree-pruning, a sum-and-send data flow starts at the leaf nodes and ends at the root node. Following the terminology used in, e.g.,~\cite{bertrand_distributed_2011_TDANSE}, this data flow is referred to as the \gls*{dansep} fusion flow, which is schematically depicted in~\figref{fig:fusionflow}.
Each node $q\in\K$ has a fusion matrix $\Pk[q]^i\in\C[M_q][Q]$ (defined later) which linearly combines its $M_q$ sensor signals into $Q$ fused signals as:

\begin{equation}\label{eq:fused_signal_zk}
    \zkd[q]^i \triangleq  (\Pk[q]^{i})^\Her\yk[q]\in\C[Q],\:\forall\:q\in\K.
\end{equation}

\noindent
In practice, iterations are spread over time such that each iteration $i$ uses a different frame of signal $\yk[q]$ to compute $\zkd[q]^i$.
Nodes receive partial in-network sums from their upstream neighbors and compute a new partial in-network sum including their own fused signals. The nodes then send the updated sum to their downstream neighbor. The partial in-network sum $\ettkq[q][l]^i$ sent from a leaf node or internal node $q\in\K\backslash\{k\}$ to its downstream neighbor $l$ is defined as:

\begin{equation}\label{eq:fusionflow}
    \ettkq[q][l]^i
    \triangleq  \zkd[q]^i + \sum_{m\in\Uk[q]^i}\ettkq[m][q]^i
    = \zkd[q]^i + \sum_{m\in\Ukb[q]^i}\zkd[m]^i.
\end{equation}

\noindent
Eventually, the root node $k$ has access to $|\Uk^i|$ partial in-network sums $\{\ettkq[l][k]^i\}_{l\in\Uk^i}$. It then constructs the so-called \gls*{dansep} observation vector $\ty^i$ as:

\begin{equation}\label{eq:yTildeRoot}
  \boxed{\ty^i \triangleq  \Big[
    \yk^\T \: \underbrace{(\ettkq[l_1][k]^{i})^\T\:\dots\:(\ettkq[l_B][k]^{i})^\T}_{(\ztk^{i})^\T}
    \Big]^\T\in\C[{\tMk^i}]}
\end{equation}

\noindent
where $\tMk^i \triangleq M_k + Q|\Uk^i|$ and $\{l_1,\dots,l_B\}=\Uk^i$. The partial in-network sums $\{\ettkq[l][k]^i\}_{l\in\Uk^i}$ are stacked in $\ztk^i\in\C[{Q|\Uk^i|}]$.

It is important to note that the observation vector in~\eqref{eq:yTildeRoot} represents a fundamental difference with \gls*{tidanse}, where the root node sums $\{\ettkq[l][k]^i\}_{l\in\Uk^i}$ into a global in-network sum, obtaining an observation vector:

\begin{equation}\label{eq:yTilde_TIDANSE}
  \begin{bmatrix}
    \yk^\T\:|\: (\sum_{l\in\Uk^i}\ettkq[l][k]^i)^\T
  \end{bmatrix}^\T = \begin{bmatrix}
    \yk^\T\:|\: (\sum_{q\neq k}\zkd[q]^i)^\T
  \end{bmatrix}^\T.
\end{equation}

\noindent
The number of \glspl*{dof} at the root node $k$ in \gls*{dansep} is thus $M_k + Q|\Uk^i|$ (cf.~\eqref{eq:yTildeRoot}), compared to $M_k + Q$ in \gls*{tidanse} (cf.~\eqref{eq:yTilde_TIDANSE}).

\subsubsection{Filter Update}\label{subsubsec:filterupdate}

After the \gls*{dansep} fusion flow, the root node $k$ estimates its target signal $\dk$ by linearly combining the elements of $\ty^i$ using a filter $\tW^{i+1}\in\C[{\tMk^i}][Q]$, obtained by solving:

\begin{equation}\label{eq:lmmseDANSEplus}
  \tW^{i+1} \triangleq  \underset{\mathbf{W}\in\C[{\tMk^i}][Q]}{\mathrm{arg\,min}}\:
  \E[
    {\left\|
        \dk - \mathbf{W}^\Her\ty^i
    \right\|^2_2}
  ].
\end{equation}

\noindent
This filter can be partitioned as:

\begin{equation}\label{eq:tW_partition}
  \tW^{i+1} =
  \Big[
    (\Wkk^{i+1})^\T \: \underbrace{(\Gkq[k][l_1]^{i+1})^\T\:\dots\:(\Gkq[k][l_B]^{i+1})^\T}_{(\Gtk^{i+1})^\T}
  \Big]^\T,
\end{equation}

\noindent
with $\Wkk^{i+1}\in\C[M_k][Q]$ the filter applied to the local sensor signals $\yk$ and $\Gtk^{i+1}\in\C[Q|\Uk^i|][Q]$ the filter applied to the partial in-network sums $\ztk^{i}$. The $\Gtk^{i+1}$ matrix is itself composed of $|\Uk^i|$ submatrices $\Gkq[k][l]^{i+1}\in\C[Q][Q],\:\forall\:l\in\Uk^i$, where $\Gkq[k][l]^{i+1}$ is applied to $\ettkq[l][k]^i$.
The solution to~\eqref{eq:lmmseDANSEplus} is an \gls*{mwf} as:

\begin{equation}\label{eq:DANSEp_mwf}
  \tW^{i+1} = \left(\Ryyt^i\right)^{-1}\Rydt^i,
\end{equation}

\noindent
with $\Ryyt^i \triangleq  \E[{\ty^i(\ty^{i})^\Her}]$ and $\Rydt^i \triangleq  \E[{\ty^i\dk^\Her}]$.
As in the centralized case, the \gls*{scm} $\Rydt^i$ can be estimated from the available signals, e.g., exploiting on-off characteristics of the desired speech signal via a \gls*{vad}. Since the signal model from~\eqref{eq:signal_model} can be used to separate $\ty^i$ into a desired signal and noise component as $\ts^i + \tn^i$, it holds that:

\begin{equation}\label{eq:Ryd_eq_Ryy_m_Rnn}
  \Rydt^i = \left(\Ryyt^i - \Rnnt^i\right)\tE^i,
\end{equation}

\noindent
with the node-specific noise \gls*{scm} $\Rnnt^i \triangleq  \E[{\tn^i(\tn^{i})^\Her}]$ and the selection matrix $\tE^i \triangleq  [\Ekk^\T \:|\: \mathbf{0}_{Q|\Uk^i|\times Q}^\T]^\T$. The estimation of $\Ryyt^i$ and $\Rnnt^i$ is discussed in detail in~\secref{subsec:implementation_aspects}.
The target signal estimate at node $k$ is then obtained as:

\begin{equation}\label{eq:dansep_dhat}
  \dhatk^{i+1} \triangleq  (\tW^{i+1})^\Her\ty^i.
\end{equation}

\noindent
Note that all nodes in the \gls*{wasn} have their own filter $\Wkk[q]^{i}$ and use it to compute their own fusion matrix $\Pk[q]^i$ (explicitly defined later). The non-root nodes $q\in\K\backslash\{k\}$ do not update their filter, i.e.:

\begin{equation}\label{eq:nonupdatingnode}
  \Wkk[q]^{i+1} = \Wkk[q]^i,\:\forall\:q\in\K\backslash\{k\}.
\end{equation}

\subsubsection{Diffusion Flow}\label{subsubsec:diffusionflow}

\begin{figure}[h]
  \centering
  \includegraphics[width=\columnwidth,trim={0 0 0 0},clip=false]{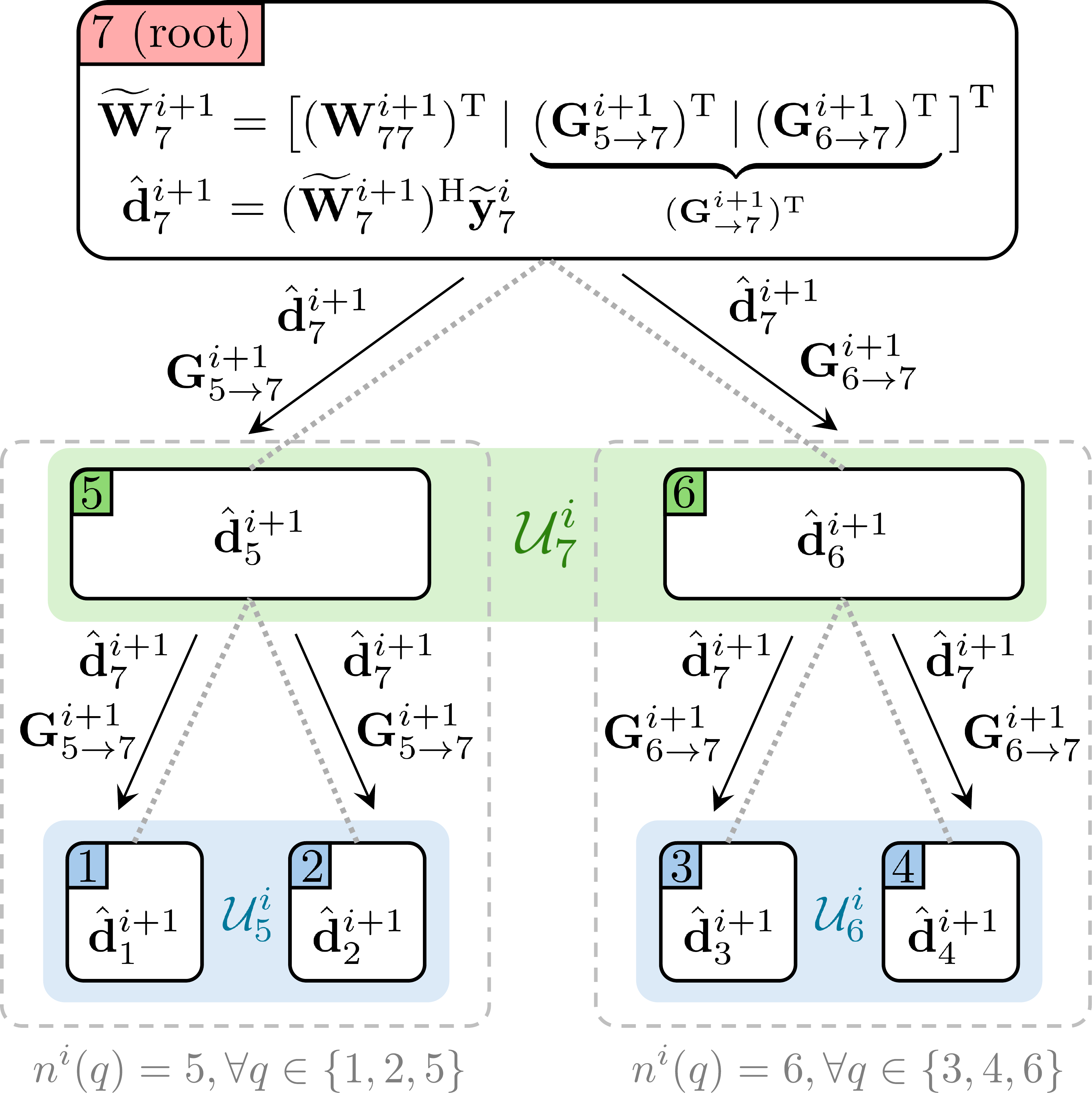}
  \caption{Schematic representation of the \gls*{dansep} diffusion flow mechanism in a topology-unconstrained \gls*{wasn} with $K=7$ nodes, pruned to a tree with root node index $k=7$. The filter $\tW^{i+1}$ is computed at the root node. The filter $\Gkq[k][l]^{i+1}$ is the part of $\tW^{i+1}$ applied to the partial in-network sum $\ettkq[l][k]^i$. It is sent through the branch starting at node $l$, and this is done $\forall l\in\Uk^i$. The index $n^i(q)$, necessary to define the fusion matrices via~\eqref{eq:fusionrule}, is shown for all $q\in\K\backslash\{k\}$ at the bottom of the figure.}
  \label{fig:diffusionflow}
\end{figure}

At the end of the filter update, the root node $k$ has access to its target signal estimate $\dhatk^{i+1}$ in~\eqref{eq:dansep_dhat} and the filter $\tW^{i+1}$ in~\eqref{eq:DANSEp_mwf}. It then shares this information with all other nodes in the \gls*{wasn}, allowing them to compute their own target signal estimates. This is done via a diffusion flow mechanism, which is similar to the fusion flow but in the opposite direction, i.e., from root node $k$ towards leaf nodes. The diffusion flow is schematically depicted in~\figref{fig:diffusionflow}.

First, node $k$ transmits $\dhatk^{i+1}$ to all its neighbors and transmits $\Gkq[k][l]^{i+1}$ to node $l$, $\forall\:l\in\Uk^i$. All nodes $l\in\Uk^i$ then forward $\dhatk^{i+1}$ and $\Gkq[k][l]^{i+1}$ to their own upstream neighbors. This is repeated until all nodes in the \gls*{wasn} have access to $\dhatk^{i+1}$ and all nodes in the $l$-th tree branch (i.e., in $\Ukb[l]^i\cup\{l\}$) have access to $\Gkq[k][l]^{i+1}$, $\forall\:l\in\Uk^i$. Although the diffusion flow requires the transmission of $|\Uk^i|$ additional $Q\times Q$ filters across the \gls*{wasn} per iteration compared to \gls*{tidanse}, the corresponding increase in communication bandwidth usage is negligible compared to the necessary transmission of batches of observations of $\{\ettkq[l][k]^i\}_{l\in\Uk^i}$ to estimate the \glspl*{scm} in~\eqref{eq:DANSEp_mwf} during the fusion flow.

For any $l\in\Uk^i$, the $Q\times Q$ filter $\Gkq[k][l]^{i+1}$ is applied by node $k$ to the partial in-network sum $\ettkq[l][k]^i$ and thus to every element of that sum, i.e., to $\zkd[q]^i,\:\forall\:q\in\Ukb[l]^i\cup l$. Since $\zkd[q]^i = (\Pk[q]^{i})^\Her\yk[q]$ according to~\eqref{eq:fused_signal_zk}, we can equivalently multiply the fusion matrix $\Pk[q]^i$ at node $q$ by the filter $\Gkq[k][l]^{i+1}$. This results in the fusion matrix definition at iteration $i+1$:

\begin{equation}\label{eq:PkDef}
  \Pk[q]^{i+1} \triangleq  \Wkk[q]^{i+1}\Tk[q]^{i+1},\:\forall\:q\in\K,
\end{equation}

\noindent
with $\Wkk[q]^{i+1}$ defined in~\eqref{eq:tW_partition} and~\eqref{eq:nonupdatingnode}. The transformation matrix $\Tk[q]^{i+1}$ is recursively defined as:

\begin{equation}\label{eq:fusionrule}
  \Tk[q]^{i+1} \triangleq  \begin{cases}
    \Tk[q]^{i}\Gkq[k][n^i(q)]^{i+1}\:&\forall\:q\in\K\backslash\{k\}\\
    \I[Q]\:&\text{for $q=k$}
  \end{cases},
\end{equation}

\noindent
where node $n^i(q)$ is the child node of node $k$ that belongs to the same branch as node $q$, i.e., $n^i(q)\in\Uk^i$ and $q\in\Ukb[n^i(q)]^i$. 
\change{
  Note that $\Tk[q]^{0}=\I[Q]\fa q\in\K$.}

Finally, node $q$ estimates its own target signal as:

\begin{align}\label{eq:desSigEst}
  \boxed{\dhatk[q]^{i+1} = \left(\Tk[q]^{i+1}\right)^{-\Her}\dhatk^{i+1},\:\forall\:q\in\K\backslash\{k\}.}
\end{align}

\noindent
A formal justification for~\eqref{eq:desSigEst} is provided in Appendix A.
Note that~\eqref{eq:desSigEst} also holds for the root node since $\Tk^{i+1} = \I[Q]$, see~\eqref{eq:fusionrule}.

In an \gls*{fc} \gls*{wasn}, \gls*{dansep} can include the \gls*{danse} formulation as a particular case if, in each iteration, the \gls*{fc} network is pruned to a star topology with the updating node as hub. In that case,~\eqref{eq:yTildeRoot} becomes equivalent to the \gls*{danse} observation vector~\cite{bertrand_distributed_2010}. All non-root nodes are then leaf nodes and thus $\ettkq[q][k]^i = \zkd[q]^i,\:\forall\:q\in\K\backslash\{k\}$, giving:

\begin{equation}\label{eq:obsvectFC}
  \widetilde{\mathbf{y}}_{k,\mathrm{FC}}^i =  \begin{bmatrix}
    \yk^\T \: (\zkd[1]^{i})^\T \:\dots\: (\zkd[k-1]^{i})^\T \:\:(\zkd[k+1]^{i})^\T \:\dots\:(\zkd[K]^{i})^\T
  \end{bmatrix}^\T.
\end{equation}

\noindent
The \gls*{dansep} algorithm is, however, advantageous in terms of communication bandwidth with respect to \gls*{danse} in \gls*{fc} \glspl*{wasn}, since \gls*{dansep} does not require the broadcasting of fused signals from all nodes in the \gls*{wasn} to every other node at each iteration. This aspect is discussed in~\secref{subsec:implementation_aspects}.

\subsection{Convergence and Optimality}\label{subsec:convergence}

The expression for the target signal estimate in~\eqref{eq:dansep_dhat} at the root node in \gls*{dansep} can be re-written using~\eqref{eq:fusionflow},~\eqref{eq:yTildeRoot},~\eqref{eq:tW_partition} and~\eqref{eq:PkDef} (the latter considering~\eqref{eq:fusionrule} in the case $q=k$) as:

\begin{align}
  \dhatk^{i+1} &= (\Wkk^{i+1})^\Her\yk + \sum_{l\in\Uk^i}(\Gkq[k][l]^{i+1})^\Her\ettkq[l][k]^i\\
  &= (\Pk^{i+1})^\Her\yk + \sum_{q\in\K\backslash\{k\}}(\Gkq[k][n^i(q)]^{i+1})^\Her(\Pk[q]^{i})^\Her\yk[q]\label{eq:rewrite_dk_eq4}\\
  &= \sum_{q\in\K}(\Pk[q]^{i+1})^\Her\yk[q].\label{eq:dansep_dhat_expanded}
\end{align}

\noindent
The step from~\eqref{eq:rewrite_dk_eq4} to~\eqref{eq:dansep_dhat_expanded} holds through~\eqref{eq:PkDef} and~\eqref{eq:fusionrule} for $q\in\K\backslash\{k\}$.
Equation~\eqref{eq:dansep_dhat_expanded} means that, at the end of each iteration, the fusion matrices are such that the sum of all fused signals corresponds to the target signal estimate of the updating node.
Substituting~\eqref{eq:dansep_dhat_expanded} into~\eqref{eq:desSigEst} gives the target signal estimate at any non-root node $q\in\K$:
\begin{equation}\label{eq:desSigEst_expanded}
  \dhatk[q]^{i+1} = \left(\Tk[q]^{i+1}\right)^{-\Her}\sum_{m\in\K}(\Pk[m]^{i+1})^\Her\yk[m],\:\forall\:q\in\K.
\end{equation}

\noindent
Based on~\eqref{eq:PkDef} and~\eqref{eq:desSigEst_expanded}, a network-wide \gls*{dansep} filter can be defined such that $(\Wk[q]^{i+1})^\Her\yk[] = \dhatk[q]^{i+1}$, $\forall\:q\in\K$, and can be written as:

\begin{equation}\label{eq:nw_filter_dansep}
  \Wk[q]^{i+1} \triangleq
  \begin{bmatrix}
      \Pk[1]^{i+1}\left(\Tk[q]^{i+1}\right)^{-1}\\
      \vdots\\
      \Pk[q]^{i+1}\left(\Tk[q]^{i+1}\right)^{-1}\\
      \vdots\\
      \Pk[K]^{i+1}\left(\Tk[q]^{i+1}\right)^{-1}\\
  \end{bmatrix}
  =
  \begin{bmatrix}
    \Wkk[1]^{i+1}\Tk[1]^{i+1}\left(\Tk[q]^{i+1}\right)^{-1}\\
    \vdots\\
    \Wkk[q]^{i+1}\\
    \vdots\\
    \Wkk[K]^{i+1}\Tk[K]^{i+1}\left(\Tk[q]^{i+1}\right)^{-1}\\
  \end{bmatrix}.
\end{equation}

\textit{Theorem 1:} Assuming $Q=S$ and $\Psibkov[q]$ full-rank, $\forall\:q\in\K$, in the signal model~\eqref{eq:sk_to_slat} and~\eqref{eq:dk_to_sk}, the solution space defined by the parametrization~\eqref{eq:nw_filter_dansep} includes the centralized optimal filters $\hWk[q]=\left(\Ryy\right)^{-1}\Rydk[q]$ as in~\eqref{eq:centr_mwf}$\fa q\in\K$.

\textit{Proof:} see Appendix B.

\textit{Theorem 2:} Assuming $Q=S$ and $\Psibkov[q]$ full-rank, $\forall\:q\in\K$, in the signal model~\eqref{eq:sk_to_slat} and~\eqref{eq:dk_to_sk}, the \gls*{dansep} algorithm in node $q$ converges to the centralized optimal filter $\hWk[q]=\left(\Ryy\right)^{-1}\Rydk[q]$ in~\eqref{eq:centr_mwf}$\fa q\in\K$.

\textit{Proof:} see Appendix C.

\textbf{Intuition for proof of Theorem 2:}
The proof of Theorem 2 is divided into several steps. In a first step, it is shown that the trajectory of \gls*{dansep} across iterations is not affected when, at a given iteration $i$ (and only at $i$), the updating node estimates the target signal of another node instead of its own. This fact can be intuitively understood by realizing that any linear transformation applied to $\dk$ at iteration $i$ in the optimization problem solved by the root node (i.e.,~\eqref{eq:lmmseDANSEplus}) will be effectively applied to all fused signals at the next iteration $i+1$ (and hence to all partial in-network sums). This transformation of the partial in-network sums can be compensated for by the next updating node (at $i+1$), when it solves the \gls*{lmmse} problem aiming at its own target signal. Combining this with the fact that all target signals in the network are linearly related to each other (cf.~\eqref{eq:dk_to_sk}), it can be concluded that letting node $k$ estimate $\dk[q]$ (with $q\neq k$) during one iteration does not affect the trajectory of \gls*{dansep}.

In a second step, this conclusion is extended to a modified version of \gls*{dansep} where all nodes in a full update cycle (i.e., $K$ consecutive iterations) aim at estimating the target signal of one fixed node $\nu\in\K$. This allows to prove convergence of \gls*{dansep} to a lower bound. In a third and final step, the optimality of this bound is proven by showing that this modified \gls*{dansep}, that does not affect the trajectory of the unmodified \gls*{dansep}, corresponds to a block coordinate descent algorithm which is known to converge to the global optimum for convex problems (such as~\eqref{eq:lmmseDANSEplus})~\cite{bertsekasParallelDistributedComputation2015}. Indeed, through one full update cycle where the target of all nodes is $\dk[\nu]$, the nodes optimize their own filter one at a time while keeping filters of non-updating nodes fixed, which is indeed equivalent to block coordinate descent.

\subsection{Tree-Pruning Strategy}\label{subsec:pruning_strategy_discussion}

The formulation of the \gls*{dansep} algorithm does not depend on a particular tree topology, such that the tree-pruning strategy may be freely chosen. \change{
The reader is reminded that an undirected \gls*{wasn} is assumed here, i.e., signals can be exchanged in both ways between any two connected nodes. The following discussion may be extended to directed graphs by considering edge directions during the pruning process, although this is not explicitly treated here.}
A typical choice is \glspl*{mst}, as they minimize the amount of energy required to transfer data through the tree by minimizing the total distance between connected nodes.
Although this is generally a popular option~\cite{szurley_topology-independent_2017}, pruning the \gls*{wasn} such that $|\Uk^i|$ is maximized at any iteration $i$ shows advantages in the context of \gls*{dansep}. Indeed, it was shown in~\cite{szurley_topology-independent_2017} that the convergence speed of \gls*{danse}-like algorithms increases with the number of available \glspl*{dof} that an updating node possesses when solving its \gls*{lmmse} problem. This explains the slow convergence speed of \gls*{tidanse}, where the updating node only possesses $M_k+Q$ \glspl*{dof} compared to $M_k + Q(K-1)$ in \gls*{danse} for \gls*{fc} \glspl*{wasn}. In constract, the size of the \gls*{dansep} observation vector at the root node~\eqref{eq:yTildeRoot} is $\tMk^i = M_k + Q|\Uk^i|$. This implies that maximizing $|\Uk^i|$ at each iteration maximizes the number of \glspl*{dof} available at the root node which, in turns, maximizes the potential for a fast \gls*{dansep} convergence.

We propose a strategy to prune a topology-unconstrained \gls*{wasn} to a tree with a maximum $|\Uk^i|,\:\forall\:i$, and with a minimum total edge cost, which may be defined, e.g., as the Euclidean distance between nodes. This is essentially a constrained version of the Kruskal's \gls*{mst}-finding algorithm~\cite{kruskal1956shortest}.
The proposed tree-pruning strategy is termed \gls*{mmut} and is explicitly defined in~\algref{alg:mmut}. It considers the fact that the maximum number of connections that root node $k$ can have after pruning is the number of neighbors that $k$ has in the original non-pruned \gls*{wasn}.
Note that applying \gls*{mmut} pruning to an \gls*{fc} \gls*{wasn} will result in a star topology where all non-root nodes are directly connected to the root.

\begin{algorithm}
  \caption{\gls*{mmut} pruning}\label{alg:mmut}
  \begin{algorithmic}[1]
    \STATE Let $\G$ denote the graph corresponding to a non-pruned topology-unconstrained \gls*{wasn} and $k$ the root node.
    \STATE Initialize tree $\Tree(k)$ with all edges of $\G$ connected to $k$.\label{algstep:mmut_step2}
    \STATE Sort the remaining edges by increasing weight (e.g., Euclidean distance between the two corresponding nodes) and list them accordingly in $\mathbf{e}_{\backslash k}$.\label{algstep:mmut_step3_sorting}
    \FOR{$e$ in $\mathbf{e}_{\backslash k}$ (in order of increasing weight)}
      \IF{$e\notin \Tree(k)$ \textbf{and} there is no path in $\Tree(k)$ between the nodes linked by $e$}
        \STATE Add $e$ to $\Tree(k)$.
      \ENDIF
    \ENDFOR\label{algstep:mmut_stepFinal}
  \end{algorithmic}
\end{algorithm}

\textbf{Remark:} Since step~\ref{algstep:mmut_step3_sorting} (edge sorting) dominates in~\algref{alg:mmut}, \gls*{mmut} pruning has the same computational complexity as Kruskal's algorithm~\cite{kruskal1956shortest}, i.e., $\mathcal{O}(E\log E)$ with $E$ the number of edges in the non-pruned \gls*{wasn}. This complexity is negligible compared to the \gls*{dansep} algorithm itself, which is dominated by the inversion of the $\tMk^i\times\tMk^i$ matrix in~\eqref{eq:DANSEp_mwf}, as described in~\secref{subsec:computational_complexity_analysis}. Furthermore, as long as the non-pruned \gls*{wasn} topology remains the same between two iterations with the same root node, the topology obtained from a previous pruning can be stored and re-used, thereby saving computations.

\subsection{Computational Complexity Analysis}\label{subsec:computational_complexity_analysis}

The most computationally costly step at any given iteration $i$ of the \gls*{dansep} algorithm is the inversion of the $\tMk^i\times\tMk^i$ \gls*{scm} $\Ryyt^i$ at the updating (root) node $k$ to compute $\tW^{i+1}$ via~\eqref{eq:DANSEp_mwf}. Each non-updating node also must invert a $Q\times Q$ matrix to compute its target signal estimate using~\eqref{eq:desSigEst}.
It follows that the total asymptotic computational complexity of the \gls*{dansep} algorithm at iteration $i$ is:

\begin{align}
  \mathrm{CC}_{\text{\gls*{dansep}}} &\triangleq O\left(
      \left(
        M_k + Q|\Uk^i|
      \right)^3 + (K-1)Q^3
    \right).
\end{align}

\noindent
This complexity is greater than that of \gls*{tidanse}, $\mathrm{CC}_{\text{\gls*{tidanse}}} \triangleq O(
  (
    M_k + Q
  )^3
)$~\cite{szurley_topology-independent_2017} and may be significantly lower than that of \gls*{danse} in \gls*{fc} \glspl*{wasn} $\mathrm{CC}_{\text{\gls*{danse}}} \triangleq O\left(
  \left(
    M_k + Q(K-1)
  \right)^3
\right)$~\cite{szurley_topology-independent_2017}, as the latter scales cubically with $K$. The \gls*{mmut} pruning strategy proposed in~\secref{subsec:pruning_strategy_discussion} allows to maximize $|\Uk^i|$ at each iteration $i$, which, although beneficial for convergence speed, increases the computational complexity of \gls*{dansep}. This underlines a trade-off to consider when defining the tree-pruning strategy.

Both \gls*{dansep} and the solution proposed in~\cite{musluoglu_distributed_2023} use partial in-network sums in a tree-pruned \gls*{wasn}. 
The solution from~\cite{musluoglu_distributed_2023} requires every non-updating node to solve its own optimization problem at each iteration, which involves the inversion of a $Q\times Q$ matrix in the case of \gls*{lmmse}-based estimation. This leads to the same total asymptotic computational cost as \gls*{dansep}, i.e., $\mathrm{CC}_{\text{\cite{musluoglu_distributed_2023}}} = \mathrm{CC}_{\text{\gls*{dansep}}}$. However, the accurate estimation of the \glspl*{scm} necessary to solve the optimization problems at non-updating nodes in~\cite{musluoglu_distributed_2023} may require a large number of signal observations, and can only be performed once the target signal at the root node is available. This limits the practical convergence speed of that solution. In contrast, non-updating nodes in \gls*{dansep} can readily compute their target signal estimate via~\eqref{eq:desSigEst} once an update has been carried out at the root node.

\subsection{Communication Complexity Analysis}\label{subsec:communication_complexity_analysis}

During the \gls*{dansep} fusion flow at any iteration $i$ (as defined in~\secref{subsec:dansep_def}), the total number of signals exchanged through the \gls*{wasn} is $Q(K-1)$, i.e., the dimension of the in-network sums multiplied by the number of communication links after tree-pruning.
During the diffusion flow, the root node $k$ floods $\dhatk^{i+1}$ through the \gls*{wasn} such that the number of signals exchanged is also $Q(K-1)$. In total, one iteration of the \gls*{dansep} algorithm corresponds to the exchange of $2Q(K-1)$ signals, which is typically much fewer than in a centralized system where $M=\sum_{q\in\K} M_q$ signals must be sent to a fusion center.

In \gls*{tidanse} too, $2Q(K-1)$ signals are exchanged at each iteration~\cite{szurley_topology-independent_2017}. In topology-unconstrained \glspl*{wasn}, \gls*{dansep} is thus nearly equivalent to \gls*{tidanse} in terms of communication bandwidth usage, since the exchange of the matrices $\Gkq[k][l]^{i+1}\fa l\in\Uk^i$ ($Q^2$ data points per matrix) is negligible in comparison to the necessary transmission of batches of signal observations to estimate the \glspl*{scm} in~\eqref{eq:DANSEp_mwf}.

However, in \gls*{fc} \glspl*{wasn}, the \gls*{danse} algorithm requires the exchange of fused signals between every node at each iteration~\cite{bertrand_distributed_2010}, leading to $KQ(K-1)$ exchanged signals per iteration, as it requires every node to build the observation vector in~\eqref{eq:obsvectFC}. If $K>2$, which is often the case in \glspl*{wasn}, \gls*{dansep} is clearly preferable to \gls*{danse} when aiming at limiting the number of exchanged signals and saving communication bandwidth.

\change{
A summary of the computational and communication complexities of the considered algorithms is provided in~\tabref{tab:comparison}.}

\begin{table}[!h]
\begin{tabular}{l|ll|}
\cline{2-3}
                                    & \multicolumn{2}{c|}{Complexity} \\ \hline
\multicolumn{1}{|c|}{Algorithm} & \multicolumn{1}{c|}{Computational} & \multicolumn{1}{c|}{Communication} \\ \hline
\multicolumn{1}{|l|}{\gls*{dansep}} & \multicolumn{1}{l|}{$O((M_k {+} Q|\Uk^i|)^3 {+} (K{-}1)Q^3)$}     & $2Q(K{-}1)$ \\ \hline
\multicolumn{1}{|l|}{\gls*{tidanse}}     & \multicolumn{1}{l|}{$O((M_k {+} Q)^3)$}     & $2Q(K{-}1)$ \\ \hline
\multicolumn{1}{|l|}{\gls*{danse}}          & \multicolumn{1}{l|}{$O((M_k {+} Q(K{-}1))^3)$}     & $KQ(K{-}1)$ \\ \hline
\multicolumn{1}{|l|}{Centralized}          & \multicolumn{1}{l|}{$O(M^3)$}     & $M$ \\ \hline
\end{tabular}
\vspace{.25em}
\caption{Computational and communication complexity for each algorithm considered.}\label{tab:comparison}
\end{table}

\subsection{Implementation Aspects}\label{subsec:implementation_aspects}

\subsubsection{SCM Estimation}\label{subsubsec:scm_estimation}

Nodes involved in \glsunset{dansep}\gls*{dansep} need the \glspl*{scm} $\Ryyt^i$ and $\Rydt^i$ to compute the \gls*{mwf}~\eqref{eq:DANSEp_mwf}. As established in~\secref{sec:prob_statement}, if the target $\dk$ and the noise-only contribution $\tn^i$ to the observation vector are uncorrelated, the matrix $\Rydt^i$ can be estimated as $\Ryyt^i - \Rnnt^i$, where the two latter \glspl*{scm} can be estimated based on observed data.

Although an exponential averaging strategy to estimate $\Ryyt^i$ and $\Rnnt^i$ can reasonably be used in the centralized case in~\eqref{eq:expavg}, it is generally unsuitable for \gls*{danse}-like algorithms (see, e.g.,~\cite[pp.9]{bertrand_distributed_2010}), \gls*{dansep} included. Indeed, since $\Ryyt^i$ and $\Rnnt^i$ are iteration-dependent, using~\eqref{eq:expavg} across iterations would result in an averaging of inconsistent signal statistics and would produce inaccurate \glspl*{scm}.

In order to retain statistical consistency in the \glspl*{scm} at play it may be advisable for all versions of \gls*{danse}, \gls*{dansep} included, to estimate the required \glspl*{scm} by averaging over frames of signals captured within a single iteration and to \textit{restart} the \gls*{scm} estimation from scratch at the beginning of the next iteration.
In practical scenarios, however, the \gls*{vad} of the desired source may only seldomly switch state. In that case, if the number of time-frames $L$ between two iterations is too small, $\Ryyt^i$ or $\Rnnt^i$ may not be properly estimated between two iterations, resulting in an inaccurate filter update. When resetting the \glspl*{scm} at every new iteration, it is instead advisable to allow $L$ to vary depending on the changes in \gls*{vad} state. One may, e.g., opt for a criterion such that $i\gets i+1$ only when both $\Ryyt^i$ and $\Rnnt^i$ have been updated a sufficient number of times.

\subsubsection{GEVD-Based Estimation}\label{subsubsec:gevddansep}

The signal estimation performance of \gls*{mwf}-based algorithms such as \gls*{dansep} can be improved when an upper bound to the number of latent target sources $S$ in~\eqref{eq:sk_to_slat} can be estimated.
Indeed, the matrix $\Rss$ in~\eqref{eq:Rydk_RssEk_final} has rank $S$, as can be seen based on~\eqref{eq:sk_to_slat} since:

\begin{equation}\label{eq:RssExpansion}
  \Rss = \E[{\mathbf{s}\mathbf{s}^\Her}] = \Psibk[]\E[{
    \mathbf{s}^\mathrm{lat}(\mathbf{s}^\mathrm{lat})^\Her
  }]\Psibk[]^\Her,
\end{equation}

\noindent
where $\Psibk[] \triangleq \begin{bmatrix}
  \Psibk[1]^\T\:\dots\:\Psibk[K]^\T
\end{bmatrix}^\T\in\C[M][S]$.
However, the estimation of $\Rss$ as $\Ryy - \Rnn$, as suggested in~\secref{sec:prob_statement}, in practice mostly results in a matrix with a rank greater than $S$ (and sometimes even not positive semidefinite) due to approximation errors. Constraining the rank of $\Rss$ to a value $R\leq S$ can be achieved using a \gls*{gevd} applied to the matrix pencil $\{\Ryy, \Rnn\}$ and yields more robust performance in scenarios where accurate estimation of $\Rss$ is challenging~\cite{doclo2002gsvd}.

\gls*{gevd}-based estimation has been applied to \gls*{danse} and \gls*{tidanse}, resulting in \glsunset{gevddanse}\gls*{gevddanse}~\cite{hassani_gevd-based_2016} and \glsunset{tigevddanse}\gls*{tigevddanse}~\cite{didier2024tigevddanse}, respectively. It can be shown that applying such a \gls*{gevd}-based rank constraint to $\Rss$ allows the corresponding unconstrained algorithm to converge even when $Q<S$~\cite{hassani_gevd-based_2016}. This is a non-negligible advantage over \gls*{danse}-like algorithms estimating $\Rss$ as $\Ryy - \Rnn$, where precise a priori knowledge of the value of $S$ is required to exactly set $Q=S$ and ensure convergence~\cite{bertrand_distributed_2010}.
Extension to \gls*{dansep} naturally follows, leading to the \glsunset{gevddansep}\mbox{\gls*{gevddansep}} algorithm. The latter is here merely proposed as a possible extension of \gls*{dansep} for improved performance and functioning when $Q\leq S$ (recall that $Q=S$ is an assumption of Theorems 1 and 2 in~\secref{subsec:convergence}). A formal convergence proof is not available, but simulation results are provided in~\secref{sec:res} to demonstrate the performance of \gls*{gevddansep}.

A \gls*{gevd} on the pencil $\{\Ryyt^i, \Rnnt^i\}$ yields:

\begin{equation}\label{eq:GEVDoutcome}
  \Ryyt^i
  = \Qkt^i \Sigk^i (\Qkt^{i})^\Her
  \:\:\:\text{and}\:\:\:
  \Rnnt^i
  = \Qkt^i (\Qkt^{i})^\Her,
\end{equation}

\noindent
where the $\tMk^i$ generalized eigenvalues are ordered from largest to smallest in $\Sigk^i = \mathrm{diag}\{
  {\sigma}_{k1}^i,...,{\sigma}_{k\tMk^i}^i
\}$ and the corresponding generalized eigenvectors are the columns of $(\Qkt^i)^{-\Her}$.
The rank of \change{
$\Rsst^i \triangleq \E[{\ts^i(\ts^i)^\Her}]$} can be constrained to a value $R$ by setting the $\tMk^i - R$ smallest generalized eigenvalues to 0, obtaining:

\begin{equation}\label{eq:RssGEVDest}
    \Rsst^i
    =
    \Ryyt^i - \Rnnt^i
    =
    \Qkt^i \mathbf{\Delta}_k^i (\Qkt^{i})^\Her,
\end{equation}

\noindent
where $\mathbf{\Delta}_k^i = \mathrm{diag}\{
    {\sigma}_{k1}^i-1, \dots, {\sigma}_{kR}^i-1, 0, \dots, 0
\}$.
This allows to rewrite the update equation at root node $k$ for \mbox{\gls*{gevddansep}} by substituting~\eqref{eq:RssGEVDest} into~\eqref{eq:DANSEp_mwf} via~\eqref{eq:Ryd_eq_Ryy_m_Rnn}:

\begin{equation}\label{eq:filterUpdateGEVDDANSEp}
    \tW^{i+1}
    =
    (\Qkt^i)^{-\Her}\tilde{\mathbf{\Lambda}}_k^i(\Qkt^{i})^\Her\tE^i,
\end{equation}

\noindent
where $\tilde{\mathbf{\Lambda}}_k^i = \mathrm{diag}\{
    1-1/{\sigma}_{k1}^i,\dots,1-1/{\sigma}_{kR}^i, 0,\dots, 0
\}$. 

To implement \gls*{gevddansep} it is sufficient to replace~\eqref{eq:DANSEp_mwf} by~\eqref{eq:filterUpdateGEVDDANSEp} in \gls*{dansep}. The other algorithmic steps remain unchanged.

\section{Simulation Results}\label{sec:res}

In this section, the simulation results are presented and discussed. In order to accurately evaluate the convergence properties of \gls*{dansep}, simulations are \change{
first} conducted using the true theoretical \glspl*{scm} computed from the steering matrices themselves. In~\secref{subsec:res_static}, static \gls*{wasn} topologies are considered, where the links between nodes remain unchanged throughout the experiment, to assess the impact of the tree-pruning strategy as introduced in~\secref{subsec:pruning_strategy_discussion}. In~\secref{subsec:res_dyn}, dynamic \gls*{wasn} topologies are introduced to showcase the robustness of \gls*{dansep} to topology-modifying events such as link failures. In~\secref{subsec:res_gevd}, the performance of the \gls*{gevddansep} algorithm introduced in~\secref{subsec:implementation_aspects} is demonstrated, also in settings where the number of latent sources ($S$) is larger than the dimension of the fused signals ($Q$).
\change{
More realistic simulations are then conducted in~\secref{subsec:res_online}, where \glspl*{scm} are estimated from the available sensor data and a \gls*{vad} as described in~\secref{subsubsec:scm_estimation}.} The performance of \gls*{dansep} is systematically compared to that of \gls*{tidanse} and of \gls*{danse} (the latter assuming an equivalent \gls*{fc} \gls*{wasn}).

\subsection{\change{Theoretical SCMs and Static Topologies}}\label{subsec:res_static}

Simulations \change{
using the true theoretical \glspl*{scm}} are conducted to evaluate the performance of the \gls*{dansep} algorithm in comparison to \gls*{danse} and \gls*{tidanse} for static \gls*{wasn} topologies.
All results are averages over $N_\mathrm{SE}=10$ randomly generated sensing environments, where each environment consists of a static non-\gls*{fc} \gls*{wasn} with $K=10$ nodes and $M_q=3$ sensors per node, $\forall\:q\in\K$, \change{
deployed in a room with reverberation time $T_{60} = 200$\,ms where} $S=1$ localized desired sound source and three localized noise sources \change{
are present}. One-dimensional signals are exchanged between nodes, i.e., $Q=1$. The desired source produces a 5\,s speech excerpt from the VCTK database~\cite{veaux2017vctk} while the noise sources are \gls*{ssn} signals, all sampled at 16\,kHz. The \gls*{snr} varies between -5\,dB and 5\,dB across nodes and environments.
Processing is conducted in the \gls*{stft}-domain with frame size $L=1024$, 50\% overlap, and a Hann-window.

\change{
The \gls*{wasn} is randomly generated on the two-dimensional slice at height $h=3$\,m of a 5\,m$\times$5\,m$\times$5\,m room}, with a minimum distance of 0.5\,m between the desired source and all sensors, and a minimum distance of 0.1\,m between any two sensors.
The desired and noise sources are randomly placed across the available space, ensuring a minimum distance of 0.5\,m between each source and all sensors. An example of simulated sensing environment is given in~\figref{fig:example_sim_wasn}.

\begin{figure}[h]
  \centering
  \begin{scriptsize}
    \input{fig4_wasn_fig.tex}
  \end{scriptsize}
  \caption{Example of randomly generated sensing environment \change{
  (top view at height $h=3$\,m in a 5\,m$\times$5\,m$\times$5\,m room)}. Nodes are represented as circles, the root node is circled in red, the desired source is a black diamond, and the noise sources are black crosses. The dashed lines represent inter-node connections pruned from the original topology-unconstrained \gls*{wasn} using \gls*{mst} pruning (left) or \gls*{mmut} pruning (right).}
  \label{fig:example_sim_wasn}
\end{figure}

To assess the convergence properties of \gls*{dansep}, the 
\glspl*{scm} required to compute~\eqref{eq:DANSEp_mwf} are computed here via their true theoretical expressions.
According to~\eqref{eq:RssExpansion}, the theoretical centralized desired signal \gls*{scm} is $\Rss = \Psibk[]\mathbf{R}_\mathbf{ss}^\mathrm{lat}\Psibk[]^\Her$ where $\mathbf{R}_\mathbf{ss}^\mathrm{lat}\in\C[S][S]$ is a diagonal matrix with the powers of the latent desired sources in $\mathbf{s}^\mathrm{lat}$ on the diagonal. Similar expressions can be obtained for $\Rnn$ and $\Ryy$. The \gls*{dansep} desired signal \gls*{scm} is then obtained by transforming $\Rss$ as $\Rsst^i = (\mathbf{C}_k^{i})^\Her\Rss\mathbf{C}_k^i$ where $\mathbf{C}_k^i\in\C[M][{\tMk^i}]$ contains an appropriate arrangement of the fusion matrices $\{\Pk[q]^i\}_{q\in\K\backslash\{k\}}$. The full expression for $\mathbf{C}_k^i$ is omitted for the sake of conciseness.

Room impulse responses are computed for each source-sensor combination using the randomized image method~\cite{desenaModelingRectangularGeometries2015}.
The impulse responses are then truncated to the frame size $L$ to fulfill the signal model assumption~\eqref{eq:sk_to_slat}. Their \gls*{dft} of length $L$ is computed and, for each frequency bin, the resulting frequency responses are used to obtain the centralized \glspl*{scm}. All other \glspl*{scm} (for \gls*{danse}, \gls*{tidanse}, and \gls*{dansep}) are computed based on the centralized \glspl*{scm} and the corresponding sets of fusion matrices.

The convergence speed of \gls*{dansep} is compared to that of \gls*{tidanse} and \gls*{danse} through \change{four} metrics, each computed at every iteration. Firstly, the \gls*{snr} at iteration $i$ is computed from the entire time-domain version $\dhatk^i[t]$ of the estimated target signal as:

\begin{equation}\label{eq:snr}
  \mathrm{SNR}^i \triangleq \frac{1}{K}
  \sum_{k=1}^K
  10\log_{10}\left(
    \frac{\sum_{t=0}^{T-1}\left\|\dhatk^i[t]\right\|_2^2}{\sum_{t=0}^{T-1}\left\|\nhatk^i[t]\right\|_2^2}
  \right),
\end{equation}

\noindent
where $\nhatk^i[t]$ is the time-domain version of the \change{
  noise component in the estimated target signal at node $k$ and iteration $i$, i.e., the noise term in} the decomposition $\dhatk^i[t]=\shatk^i[t]+\nhatk^i[t]$ (see~\eqref{eq:signal_model}), and $T$ is the number of time-domain samples (given the simulation conditions, $T=32\cdot10^4$).

Secondly, the \gls*{stoi}~\cite{taalAlgorithmIntelligibilityPrediction2011} is computed \change{
for each iteration $i$} over the \change{speech-active segments of the} time-domain version $\dhatk^i[t]$ of the estimated target signal, with the \change{corresponding speech-active segments of the} true target signal as reference.

\change{Thirdly, the \gls*{pesq} is computed for iteration over the same speech-active segments as for \gls*{stoi}, using the wideband mode~\cite{rixPerceptualEvaluationSpeech2001}.}

\change{Finally}, the \gls*{mse} between the network-wide filters\footnote{For mathematical expressions of the network-wide filters, the reader is referred to~\eqref{eq:nw_filter_dansep} for \gls*{dansep}, \cite[pp.8]{bertrand_distributed_2010} for \gls*{danse}, and~\cite[pp.6]{szurley_topology-independent_2017} for \gls*{tidanse}.} $\Wk[q]^{i}$ at iteration $i$ and the centralized filters $\hWk[q]$ from~\eqref{eq:centr_mwf} is computed. It is defined as:

\begin{equation}
  \mathrm{MSE}_{W}^i \triangleq \frac{1}{K} \sum_{q=1}^K \left\|
    \Wk[q]^{i} - \hWk[q]
  \right\|^2_\mathrm{F}.
\end{equation}

\noindent
Note that, since processing is conducted in the \gls*{stft}-domain, the filters $\Wk[q]^{i}$ and $\hWk[q]$ are frequency-dependent, i.e., they are computed for each frequency bin separately. The \gls*{mse} is thus computed for each frequency bin and averaged across all bins to obtain a single value per iteration $i$.

The convergence profile of the algorithms in terms of $\mathrm{MSE}_{W}^i$ can be expected to differ between sensing environments due to the different relative node-to-source positions.
In order to unambiguously depict the average $\mathrm{MSE}_{W}^i$ values obtained across multiple sensing environments, we plot a geometric mean on a logarithmic axis, defined as:

\begin{equation}\label{eq:geommean}
  \overline{\mathrm{MSE}}_{W}^i \triangleq \left(
    \prod_{m=1}^{N_\mathrm{SE}} \mathrm{MSE}_{W}^i(m)
  \right)^{\frac{1}{N_\mathrm{SE}}},
\end{equation}

\noindent
where $\mathrm{MSE}_{W}^i(m)$ denotes the value obtained for the $m$-th sensing environment at iteration $i$.

As discussed in~\secref{subsec:pruning_strategy_discussion}, the number of neighbors at the root node is expected to influence the convergence speed of \gls*{dansep}. For a fixed $K$, the average number of neighbors per node $|\mathcal{U}_\mathrm{avg}|$ increases with the number of existing connections between nodes, reaching a maximum value of $K{-}1$ in \gls*{fc} \glspl*{wasn}. We define the following measure of connectivity:

\begin{equation}\label{eq:connectivity}
  C\triangleq \frac{\mathbf{1}^\T\mathbf{\Lambda}\mathbf{1} - 2K}{K(K-3)},
\end{equation}

\noindent
where $\mathbf{1}$ denotes the $K$-dimensional all-ones column vector and $\mathbf{\Lambda}\in\{0,1\}^{K\times K}$ is the adjacency matrix of the \gls*{wasn}, i.e., $[\mathbf{\Lambda}]_{q,l} = 1$ if $q$ and $l{\neq} q$ are connected, 0 otherwise, and $[\mathbf{\Lambda}]_{q,q} = 0$ by convention, $\forall\:q\in\K$. In an \gls*{fc} network, $C=1$ while, in a tree topology, $C=0$. As $C$ is unambiguously bounded between 0 and 1, we prefer it over the algebraic connectivity, whose upper bound depends on the graph structure~\cite{fiedler1973algebraic}, for the sake of clarity.
To evaluate the convergence speed of \gls*{dansep} in \glspl*{wasn} with various values of $|\mathcal{U}_\mathrm{avg}|$, we randomly remove or establish transmission links between nodes of the original \gls*{wasn} (ensuring that the network remains connected) until a given value of $C$ is obtained. This process is repeated for different values of $C$.

The $\overline{\mathrm{MSE}}_{W}^i$ is computed when pruning the topology-unconstrained \gls*{wasn} at each iteration $i$ to either (i) an \gls*{mst} with root $k$ using Kruskal's algorithm with the inter-node Euclidean distance as edge weight or (ii) a maximum-$|\Uk|$ tree with root $k$ obtained via \gls*{mmut} pruning. The results for \gls*{danse} are obtained in the case where the \gls*{wasn} would be \gls*{fc}, which makes them independent of $C$. The results obtained for \gls*{tidanse} are also independent of $C$.

\begin{figure}[h]
  \centering
  \includegraphics[width=\columnwidth,trim={0 0 0 0},clip=false]{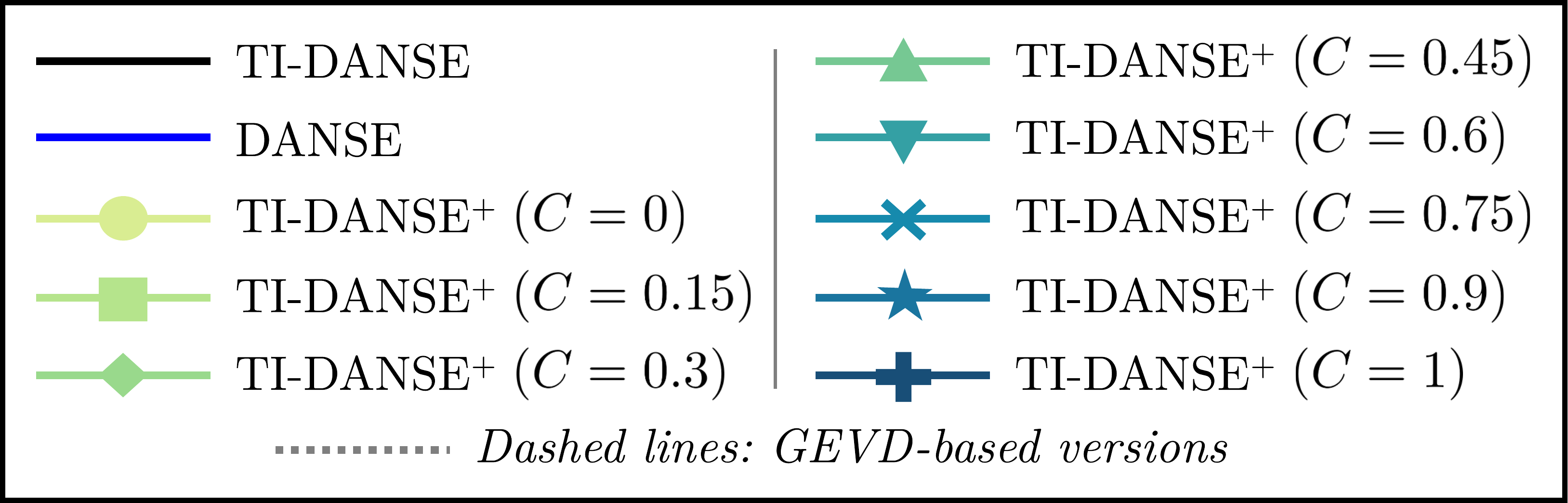}
  \caption{Line color and marker styles for the different algorithms.}
  \label{fig:legend}
\end{figure}

\begin{figure}[h]
  \centering
  \includegraphics[width=\columnwidth,trim={0 0 0 0},clip=false]{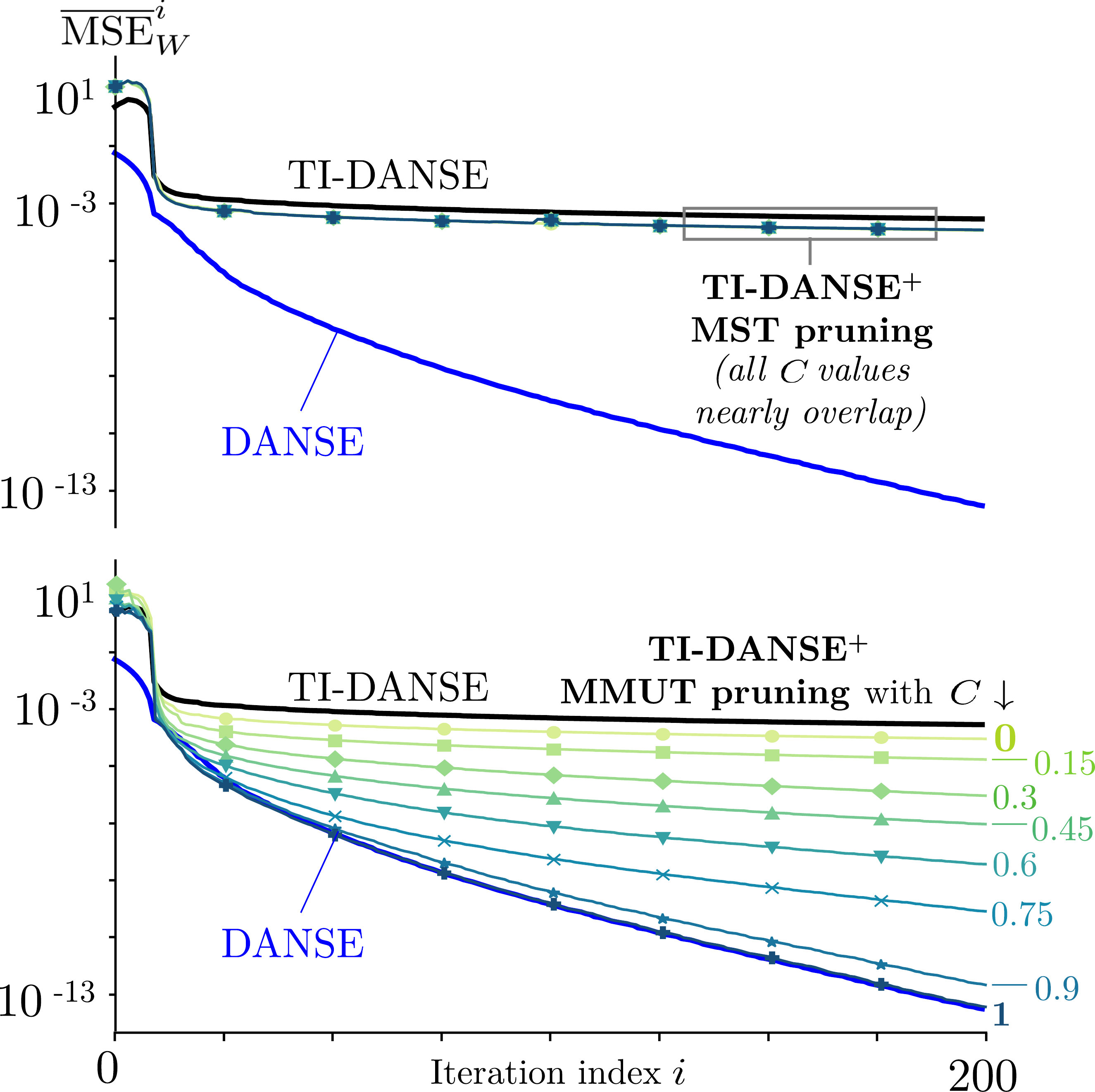}
  \caption{$\overline{\mathrm{MSE}}_{W}^i$ for \gls*{dansep} with different values of connectivity $C$, with comparison to \gls*{tidanse} and \gls*{danse}, the latter obtained as if the \gls*{wasn} was \gls*{fc}. Averages over 10 randomly generated sensing environments. The results using \gls*{mst} or \gls*{mmut} pruning are depicted in the top and bottom plots, respectively.}
  \label{fig:res_msew}
\end{figure}

\begin{figure}[h]
  \centering
  \includegraphics[width=\columnwidth,trim={0 0 0 0},clip=false]{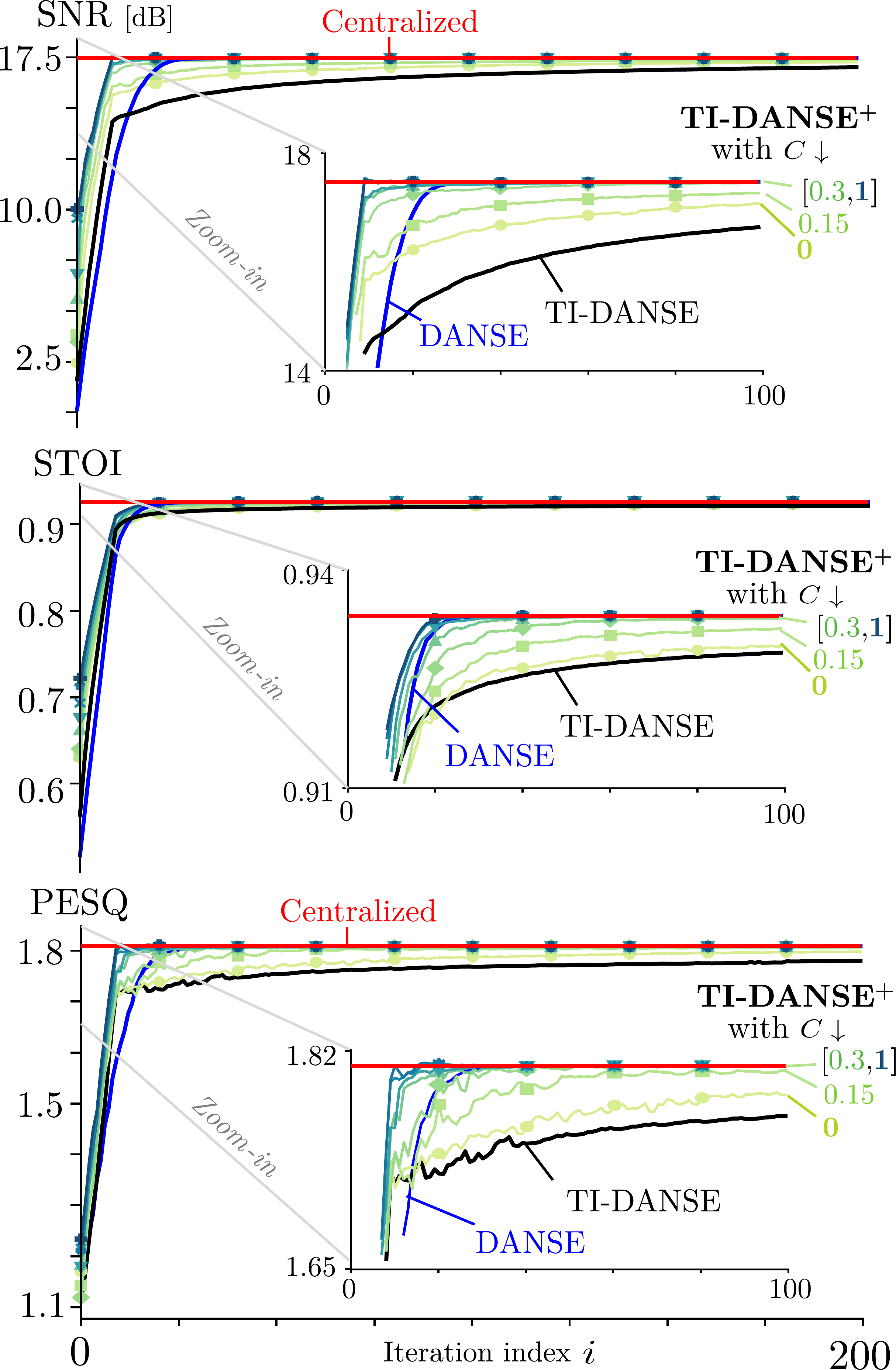}
  \caption{\change{Average \gls*{snr} (top), \gls*{stoi} (middle), and \gls*{pesq} (bottom)} over all $K=10$ nodes for \gls*{dansep} \change{with \gls*{mmut}  pruning and} different values of connectivity $C$, with comparison to \gls*{tidanse} and \gls*{danse}, the latter obtained as if the \gls*{wasn} was \gls*{fc}. Averages over 10 randomly generated sensing environments.}
  \label{fig:res_metrics}
\end{figure}

\change{
For clarity, a legend depicting the line colors and marker styles used for the different algorithms throughout this section is provided in~\figref{fig:legend}. The $\overline{\mathrm{MSE}}_{W}^i$ values for static topologies are shown in~\figref{fig:res_msew}. The \gls*{snr} and \gls*{stoi} metrics are shown, only for \gls*{mmut} pruning, in~\figref{fig:res_metrics}. 
}

Firstly, the \change{ 
$\overline{\mathrm{MSE}}_{W}^i$ plots in \figref{fig:res_msew}} underline the slow convergence of \gls*{tidanse} compared to that of \gls*{danse} in an equivalent \gls*{fc} \gls*{wasn}. Secondly, they highlight the importance of the tree-pruning strategy for \gls*{dansep}, as discussed in~\secref{subsec:pruning_strategy_discussion}. On the one hand, when pruning the topology-unconstrained \gls*{wasn} to an \gls*{mst}, the original connectivity $C$ does not play any significant role for \gls*{dansep}. On the other hand, \gls*{mmut} pruning enables \gls*{dansep} to exploit its full potential, showing a higher convergence speed as the original topology-unconstrained \gls*{wasn} becomes more connected. In fact, for $C=1$ (\gls*{fc}), the convergence speed of \gls*{dansep} matches that of \gls*{danse} when using \gls*{mmut} pruning. The benefit of using \gls*{dansep} instead of \gls*{tidanse} in non-\gls*{fc} \glspl*{wasn} is visible for any value of $C$, i.e., even in relatively sparse networks with few inter-node connections. Even with $C=0$ (i.e., in \glspl*{wasn} with a tree topology) or when pruning to an \gls*{mst}, the \gls*{dansep} algorithm reaches a lower $\overline{\mathrm{MSE}}_{W}^i$ in fewer iterations than \gls*{tidanse}. This is due to the fact that, even in a tree topology with $K>2$, at least one node has more than one neighbor and can, through \gls*{dansep}, exploit several partial in-network sums individually to update its filters.

\change{All metrics in \figref{fig:res_metrics}, obtained for \gls*{mmut} pruning only,} confirm the conclusions drawn from the $\overline{\mathrm{MSE}}_{W}^i$ plots, i.e., that \gls*{dansep} outperforms \gls*{tidanse} and that \gls*{mmut} pruning is beneficial for \gls*{dansep}. Interestingly, the \gls*{snr} \change{and \gls*{pesq} metrics show more pronounced improvements} for \gls*{dansep} with \gls*{mmut} pruning compared to \gls*{tidanse} than the \gls*{stoi} metric, which is more stable across different values of $C$. This suggests that \gls*{dansep} is particularly effective in quickly (i.e., after few iterations) improving the signal-to-noise ratio \change{and \gls*{pesq}} of the estimated target signal.

\subsection{\change{Theoretical SCMs and Dynamic Topologies}}\label{subsec:res_dyn}

\begin{figure}[h]
  \centering
  \includegraphics[width=\columnwidth,trim={0 0 0 0},clip=false]{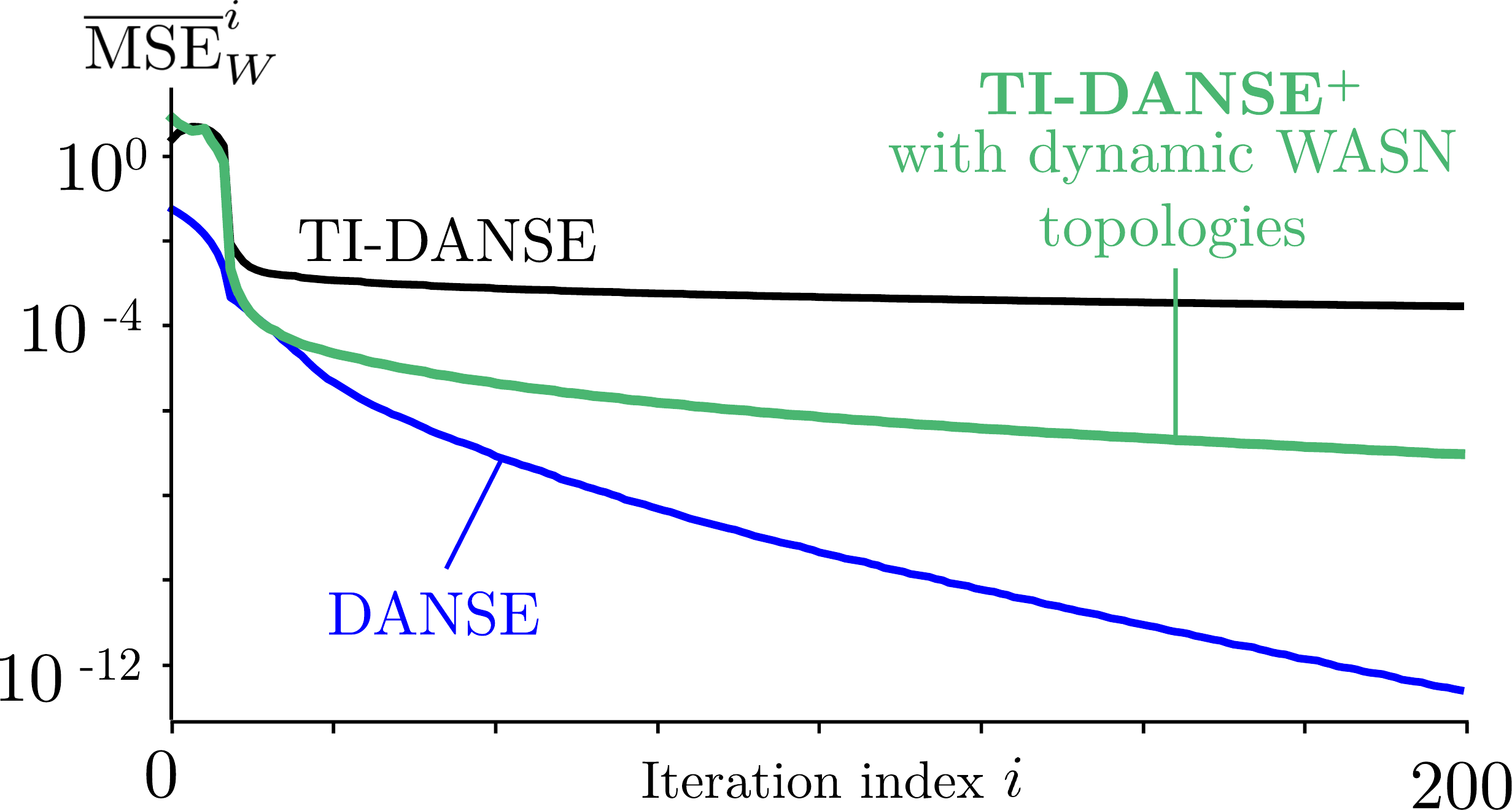}
  \caption{$\overline{\mathrm{MSE}}_{W}^i$ for \gls*{dansep} with \gls*{mmut} pruning, \gls*{tidanse}, and \gls*{danse}, the latter obtained as if the \gls*{wasn} was \gls*{fc}. Averages over 10 dynamic sensing environments with randomized link failures.}
  \label{fig:res_msew_dyn}
\end{figure}

In this section, the performance of \gls*{dansep} \change{
using the true theoretical \glspl*{scm}} is evaluated for dynamic environments to assess the robustness of \gls*{dansep} to link failures and other topology-modifying events. The same simulation parameters as in~\secref{subsec:res_static} are used. A new adjacency matrix is generated at every iteration $i$, randomly establishing or cutting the communication between nodes. It is ensured that the resulting topology remains always connected so that the centralized \gls*{mwf} stays the same. The results are presented for \gls*{mmut} pruning in~\figref{fig:res_msew_dyn}, as an average over 10 simulated sensing environments each with randomly generated adjacency matrices at every new iteration. Once again, the results obtained with \gls*{tidanse} and \gls*{danse} (assuming \gls*{fc} \glspl*{wasn} for the latter) are shown as well.
Due to the adjacency matrix randomization, the value of $C$ as defined in~\eqref{eq:connectivity} changes at every iteration. 

The results show that \gls*{dansep} still provides a substantial convergence speed improvement compared to \gls*{tidanse} and retains convergence towards the centralized \gls*{mwf} despite the link failures.

\subsection{TI-GEVD-DANSE$^+$ \change{with Theoretical SCMs}}\label{subsec:res_gevd}

\begin{figure}[h]
  \centering
  \includegraphics[width=\columnwidth,trim={0 0 0 0},clip=false]{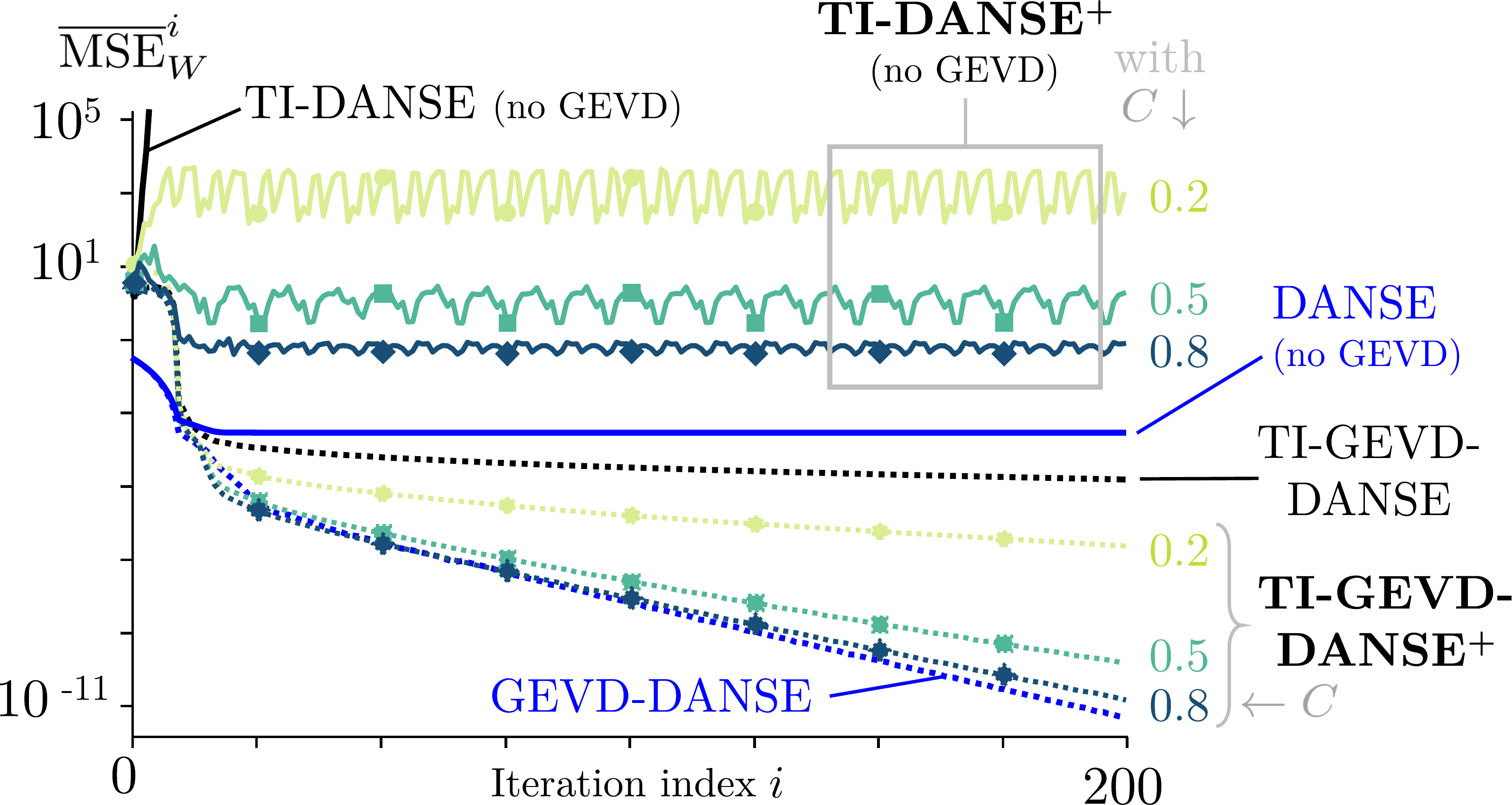}
  \caption{$\overline{\mathrm{MSE}}_{W}^i$ for \gls*{dansep} with \gls*{mmut} pruning in comparison to \gls*{tidanse} and \gls*{danse}, the latter obtained as if the \gls*{wasn} was \gls*{fc}, in a scenario with $Q<S$, with or without \gls*{gevd}-based filter updates. Averages over $N_\mathrm{SE}=10$ sensing environments.}
  \label{fig:res_msew_gevd}
\end{figure}

In order to showcase the potential of the \gls*{gevddansep} algorithm introduced in~\secref{subsec:implementation_aspects}, we repeat part of the simulations \change{
using the true theoretical \glspl*{scm}} from~\secref{subsec:res_static} while changing the number of desired sources to $S=3$. The setting $Q=1$ is kept, such that $Q<S$, which violates a core assumption of Theorem 2, i.e., theoretical convergence of \gls*{dansep} is not guaranteed anymore. The performance of \gls*{dansep}, \gls*{danse}, and \gls*{tidanse} in terms of $\overline{\mathrm{MSE}}_{W}^i$ is compared to that of \gls*{gevddansep} (using~\eqref{eq:filterUpdateGEVDDANSEp}), \gls*{gevddanse}~\cite{hassani_gevd-based_2016}, and \gls*{tigevddanse}~\cite{didier2024tigevddanse}, respectively, all with a \gls*{gevd} rank $R=1$. Again, \gls*{mmut} pruning is used at every iteration. The results, showcased in~\figref{fig:res_msew_gevd}, confirm that the non-\gls*{gevd}-based algorithms do not converge towards the centralized \gls*{mwf} in such $Q<S$ setting. In contrast, the \gls*{gevd}-based algorithms do converge, highlighting the robustness provided by constraining the rank of the desired-signal \gls*{scm}.

\subsection{TI-GEVD-DANSE$^+$ with Estimated SCMs}\label{subsec:res_online}

The \gls*{gevddansep} algorithm with \gls*{mmut} pruning is evaluated in simulations where \glspl*{scm} are \textit{estimated using sensor data in a realistic acoustic environment}, instead of using true theoretical \glspl*{scm} obtained from the steering matrices as in the previous experiments. The performance of \gls*{gevddansep} is compared to the respective performances of \gls*{gevddanse} and \gls*{tigevddanse}.

A static acoustic scenario is simulated using \textit{measured} \glspl*{rir} from the ISOBEL dataset~\cite{Kristoffersen2021_ISOBEL} (corresponding to a room with measured $T_{20}$ of 0.46\,s below 316\,Hz). The $K=5$ node positions are randomly generated and, for each node $k\in\K$, the $M_k=3$ closest \gls*{rir} measurement positions are selected as sensors for that node, ensuring that no two nodes share the same sensor positions. Since \glspl*{rir} are only available for two source positions in the dataset, we assign one to be the desired (speech) source and the other a noise source. The signals are simulated by convolving the latent source signals (speech from the VCTK database~\cite{veaux2017vctk} and \gls*{ssn}) with the corresponding \glspl*{rir}. Sensor noise is also added as in~\secref{subsec:res_static}.

The \glspl*{scm} are estimated at each node using time-averages over small chunks of available signals, for each algorithmic iteration (i.e., each index $i$) separately. This effectively corresponds to the \gls*{scm} resetting strategy motivated in~\secref{subsec:implementation_aspects}.
An energy-based node-specific \gls*{vad} computed on the noisy signal captured by the first sensor of each node, as may be used in a real-world setting, is used to isolate speech+noise from noise-only periods.
Taking the \gls*{dansep} speech+noise \gls*{scm} at node $k$ as example, it is approximated as $\Ryyt^i \approx 1/L_\mathrm{avg}\sum_{l\in\mathcal{L}}\ty[k]^i[l](\ty[k]^i[l])^\Her$, where $\ty[k]^i[l]$ is the observation vector at node $k$ and frame $l$, and $\mathcal{L}$ is the set of the $L_\mathrm{avg}$ most recently recorded \gls*{stft} frame indices where \gls*{vad}=1 (speech+noise). An analogous expression is used for the noise-only \glspl*{scm}, considering the \gls*{vad}=0 (noise-only) frames.
The performance of each algorithm is evaluated through the \gls*{snr}, \gls*{stoi}, and \gls*{pesq} computed at every iteration $i$ over a 10\,s chunk of estimated target signal. This estimate is computed, for each algorithm and each node, by applying the network-wide version of the filters\footnote{The use of a \gls*{gevd} in the filter update equation (as in~\eqref{eq:filterUpdateGEVDDANSEp}) does not modify the network-wide filter expressions.} obtained at iteration $i$ to 10\,s of centralized noisy signal vector $\yk[]$.

\begin{figure}[h]
  \centering
  \includegraphics[width=\columnwidth,trim={0 0 0 0},clip=false]{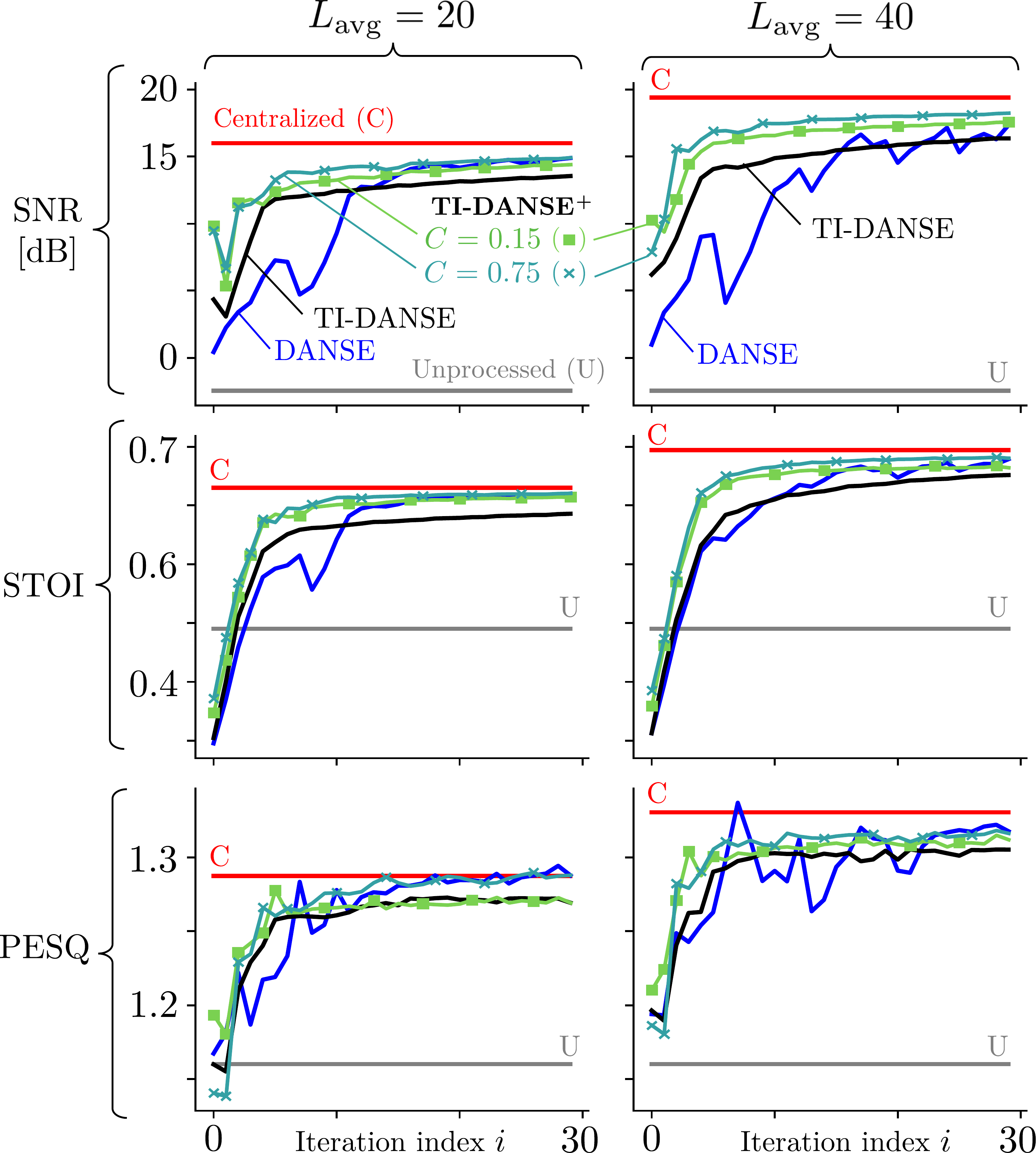}
  \caption{\change{\gls*{snr} (top), \gls*{stoi} (middle), and \gls*{pesq} (bottom) for \gls*{gevddansep} with \gls*{mmut} pruning in comparison to \gls*{gevddanse}, \gls*{tigevddanse}, and the centralized \gls*{gevdmwf}. \glspl*{scm} required to obtain the results were estimated using time-averages over chunks of data simulated using measured \glspl*{rir}. Results are shown for two chunk length of $L_\mathrm{avg}=20$ (left) and 40 (right).}}
  \label{fig:res_online}
\end{figure}

The results are presented in~\figref{fig:res_online} as averages over 10 randomly generated sensing environments, for \gls*{wasn} connectivities $C\in\{0.15, 0.75\}$ and batch sizes $L_\mathrm{avg}\in\{20,40,100\}$ frames (corresponding to a minimum of 0.64\,s, 1.28\,s, and 3.2\,s, respectively, required to estimate each \gls*{scm} (for each iteration) when using a $L=1024$-point \gls*{stft} with 50\% window overlap).
The performance of a centralized \gls*{gevdmwf} obtained using a single batch-estimate of the centralized \glspl*{scm} over the entire 10\,s signal segment is also shown as reference. 

The results in~\figref{fig:res_online} show that \gls*{gevddansep} generally outperforms \gls*{tigevddanse} in terms of \gls*{snr}, \gls*{stoi}, and \gls*{pesq}, even for both network connectivities and batch sizes tested. This gain is less pronounced in terms of \gls*{pesq} under the most challenging combination of parameters $C=0.15$ and $L_\mathrm{avg}=20$, but the performance of \gls*{tigevddanse} is still matched.
It can be noted that the relatively low values of \gls*{stoi} and \gls*{pesq} may be attributed to the challenging acoustic environment considered, with a high reverberation time and estimated \gls*{vad}.
The slow convergence of \gls*{tigevddanse} observed in~\secref{subsec:res_static} with theoretical \glspl*{scm} is visible, comparing with \gls*{gevddanse} and \gls*{gevddansep}.

In the early iterations, the performance of \gls*{gevddanse} is lesser than that of \gls*{tigevddanse}, which can be attributed to the fact that \gls*{scm} estimation may be more challenging in \gls*{gevddanse} than in \gls*{tigevddanse} due to the larger size of the observation vector. The results in~\figref{fig:res_online} correlate well with those obtained with theoretical \glspl*{scm} in~\secref{fig:res_metrics}.
As expected, increasing the chunk size $L_\mathrm{avg}$ used for \gls*{scm} estimation leads to improved performance for all algorithms, although increasing the inter-iteration time required for \gls*{scm} estimation. For \gls*{gevddansep} using \gls*{mmut} pruning, the performance gap to the centralized \gls*{gevdmwf} is further reduced for more connected \glspl*{wasn} (i.e., higher $C$).

\section{Conclusion}\label{sec:ccl}

We proposed \gls*{dansep}, a distributed \gls*{lmmse}-based algorithm for node-specific signal estimation in topology-unconstrained \glspl*{wasn}. The algorithm achieves rapid convergence toward the centralized \gls*{mwf} estimate without requiring a fully connected (\gls*{fc}) topology. To increase convergence speed, a tree-pruning strategy was introduced that maximizes the number of neighbors at the updating node.

The results demonstrate that \gls*{dansep} serves as an effective single-algorithm replacement for the combined use of \gls*{danse} and \gls*{tidanse}. It consistently outperforms \gls*{tidanse} in non-\gls*{fc} networks and matches the convergence speed of \gls*{danse} in \gls*{fc} topologies, while reducing the amount of data exchanged between nodes in that case. The choice of tree-pruning strategy is shown to be a key factor in achieving these improvements.

However, challenges remain, particularly with respect to \change{real-time} online operation.
\change{
  In particular, the fast and accurate estimation of signal statistics (i.e., \glspl*{scm}) from limited data at each iteration, although paramount to compute the \glspl*{mwf} involved in \gls*{dansep} (as well as \gls*{danse} and \gls*{tidanse}), is still difficult in practice, especially in dynamic acoustic scenarios. Potential solutions include the use of advanced \gls*{scm} estimation techniques, possibly based on deep learning, which could be explored in future work.
}

\vspace{-.7em}
\section*{Appendix}

\subsection{Target Signal Estimates at Non-Updating Nodes}\label{subapp:proof_dq}

We will here show that~\eqref{eq:desSigEst} must hold in order to guarantee optimality at all nodes upon \gls*{dansep} convergence.
Consider a particular \gls*{dansep} iteration $i$ where $k$ is the updating node. Since it holds that $\dhatk^{i+1} = \sum_{q\in\K}(\Pk[q]^{i+1})^\Her\yk[q]$ (cf.~\eqref{eq:dansep_dhat_expanded}), we have that the network-wide \gls*{dansep} filter at the updating node $k$ is simply a stacked version of the updated fusion matrices at all nodes:

\vspace{-.5em}

\begin{equation}\label{eq:nw_filt_upnode}
    \Wk^{i+1} = \begin{bmatrix}
        (\Pk[1]^{i+1})^\T\:|\:\dots\:|\:(\Pk[K]^{i+1})^\T
    \end{bmatrix}^\T.
\end{equation}

\noindent
The centralized filter from~\eqref{eq:centr_mwf} can also be partitioned as:
\vspace{-.5em}

\begin{equation}\label{eq:centr_mwf_partition}
    \hWk \triangleq \begin{bmatrix}
        \hWk[k1]^\T\:|\:\dots\:|\:\hWk[kK]^\T
    \end{bmatrix}^\T,
\end{equation}

\noindent
where $\hWk[kq]\in\C[M_{q}][Q]$ is applied to $\yk[q]\fa q\in\K$. Suppose now that optimality has been reached at iteration $i$. The target signal estimate $\dhatk[k]^{i+1}$ is then equal to the centralized estimate $\hWk[k]^\Her\yk[]$, i.e., $\Wk^{i+1}=\hWk[k]$.
Comparing~\eqref{eq:nw_filt_upnode} and~\eqref{eq:centr_mwf_partition} gives:
\vspace{-.5em}

\begin{equation}\label{eq:Pq_eq_hWkq}
    \Pk[q]^{i+1} = \hWk[kq]\fa q\in\K.
\end{equation}

\noindent
At updating node $k$,~\eqref{eq:Pq_eq_hWkq} is equivalent to $\Wkk[k]^{i+1} = \hWk[kk]$ since $\Tk^{i+1}=\mathbf{I}_Q$. Now consider the following iteration $i+1$ where $k'\neq k$ is the updating node. If optimality is maintained at $i+1$, it holds that $\dhatk[k']^{i+2}=\hWk[k']^\Her\yk[]$ and thus $\Wkk[k']^{i+2} = \hWk[k'k']$. Considering \gls*{dansep} upon convergence, optimality is maintained at all iterations after $i$. Taking the sequential updating scheme~\eqref{eq:nonupdatingnode} into account, this gives:

\begin{equation}\label{eq:Wqq_eq_hWqq}
    \exists\:n\::\:
    \forall\: j\geq n,\:
    \Wkk[q]^{j+1} = \hWk[qq]\fa q\in\K.
\end{equation}

\noindent
If $Q=S$ and $\Psibkov[q]$ is full-rank (and thus invertible), all centralized optimal filters $\hWk[q]$ share the same column space since, using~\eqref{eq:dk_to_sk} in~\eqref{eq:centr_mwf}:
\vspace{-1em}

\begin{align}
    \hWk[q] &= \left(\Ryy\right)^{-1}\Rydk[q] = \left(\Ryy\right)^{-1}\E[\mathbf{yd}_q^\Her]\\
    &= \left(\Ryy\right)^{-1}\E[\mathbf{y}(\mathbf{s}^\mathrm{lat})^\Her]\Psibkov[q]^\Her,\:\forall\:q\in\K,\\
    \Rightarrow\:&\boxed{\forall\:(q,q')\in\K^2:\hWk[q] = \hWk[q']\left(\Psibkov[q']\right)^{-\Her}\Psibkov[q]^\Her.}\label{eq:Akq}
\end{align}

\noindent
Considering \gls*{dansep} upon convergence,~\eqref{eq:Pq_eq_hWkq} can be more generally written as:
\vspace{-1em}

\begin{align}
    \exists\:n\::\:
    \forall\: j\geq n,\:
    \Pk[q]^{j+1} &= \hWk[k(j)q],\label{eq:Pq_eq_hWkjq}
\end{align}

\noindent
where $k(j)$ denotes the root node at iteration $j$. We can use condition~\eqref{eq:Pq_eq_hWkjq} to rewrite transformation matrix $\Tk[q]^{j+1}$.
First, inserting~\eqref{eq:Wqq_eq_hWqq} into~\eqref{eq:PkDef}, we find:
\vspace{-1em}

\begin{align}\label{eq:Pq_eq_hWkjq2}
    \exists\:n\::\:
    \forall\: j\geq n,\:
    \Pk[q]^{j+1} &= \hWk[qq]\Tk[q]^{j+1}.
\end{align}

\noindent
Then, according to~\eqref{eq:Akq} and~\eqref{eq:Pq_eq_hWkjq2},~\eqref{eq:Pq_eq_hWkjq} only holds if: 
\vspace{-1em}

\begin{align}\label{eq:optimality_condition_Tk}
    \exists\:n\::\:
    \forall\: j\geq n,\:
    \Tk[q]^{j+1} &= \left(\Psibkov[q]\right)^{-\Her}\Psibkov[k(j)]^\Her
    \fa q\in\K.
\end{align}

\noindent
Upon \gls*{dansep} convergence, condition~\eqref{eq:optimality_condition_Tk} can be used to express the centralized desired signal estimate $\dhatk[q]$ as a function of the target signal estimate at the root node:
\vspace{-1em}

\begin{align}
    \exists\:n\::\:
    \forall\: j\geq n,
    \dhatk[q]
    &=  \left(\hWk[k(j)]\left(\Psibkov[k(j)]\right)^{-\Her}\Psibkov[q]^\Her\right)^\Her\yk[]\\
    &= \left(\left(\Psibkov[q]\right)^{-\Her}\Psibkov[k(j)]^\Her\right)^{-\Her}\hWk[k(j)]^\Her\yk[]\\
    &= \left(\Tk[q]^{j+1}\right)^{-\Her}\dhatk[k(j)]^{j+1}
    \fa q\in\K.\label{eq:desSigEst_alt}
\end{align}

\noindent
since $\dhatk[k(j)]^{j+1} = \hWk[k(j)]^\Her\yk[]\fa j\geq n$.
Equation~\eqref{eq:desSigEst_alt} shows that, at any iteration $j$ after \gls*{dansep} convergence is reached, the optimal target signal estimate at any non-updating node $\dhatk[q]^{j+1}=\dk[q]$ can be obtained by transforming $\dhatk^{j+1}$ via $\left(\Tk[q]^{j+1}\right)^{-\Her}$. This justifies relation~\eqref{eq:desSigEst}.

\vspace{-1em}
\subsection{Proof of Theorem 1}\label{subapp:proof_th1}

By setting $\Tk[q]^{i+1} = \left(\Psibkov[q]\right)^{-\Her}\Psibkov^\Her\fa q\in\K$, where $k$ is still the root node index (consistent with the definition in~\eqref{eq:fusionrule} since $\left(\Psibkov\right)^{-\Her}\Psibkov^\Her = \mathbf{I}_Q$), it holds that:

\begin{align}\label{eq:TkTq_to_Akq}
  \Tk[q']^{i+1}\left(\Tk[q]^{i+1}\right)^{-1} &= \left(\Psibkov[q']\right)^{-\Her}\Psibkov^\Her\left(\left(\Psibkov[q]\right)^{-\Her}\Psibkov^\Her\right)^{-1}\\
  &= \left(\Psibkov[q']\right)^{-\Her}\Psibkov[q]^\Her\fa(q,q')\in\K^2.
\end{align}

\noindent
Setting $\Wkk[q]^{i+1} = \hWk[qq]\fa q\in\K$ and inserting this together with~\eqref{eq:TkTq_to_Akq} into~\eqref{eq:nw_filter_dansep}, and taking~\eqref{eq:Akq} into account, we obtain:

\begin{equation}\label{eq:showing_inclusion_in_solspace}
    \Wk[q]^{i+1}
    =
    \begin{bmatrix}
        \hWk[11]\left(\Psibkov[1]\right)^{-\Her}\Psibkov[q]^\Her\\
        \vdots\\
        \hWk[qq]\\
        \vdots\\
        \hWk[KK]\left(\Psibkov[K]\right)^{-\Her}\Psibkov[q]^\Her\\
    \end{bmatrix}
    =
    \hWk[q],
\end{equation}

\noindent
which shows that the centralized optimal filter $\hWk[q]\fa q\in\K$ is indeed included in the solution space defined by~\eqref{eq:nw_filter_dansep}.
\hfill$\blacksquare$

\subsection{Proof of Theorem 2}\label{subapp:proof_th2}
We recall here that the \gls*{dansep} is defined for sequential node updating as \gls*{danse}~\cite{bertrand_distributed_2010} and \gls*{tidanse}~\cite{szurley_topology-independent_2017}, i.e., nodes update their local filters in a round-robin fashion, one after the other, at each iteration. In the following, a \textit{full update cycle} is defined as $K$ consecutive iterations. Through a full update cycle, each node updates its local filters once.

In Step 1, we will consider a first modified version of \gls*{dansep} where the updating node at a given iteration estimates the target signal of another node instead of its own. We will show that this does not affect the \gls*{dansep} trajectory in later iterations. In Step 2, we will consider a second modifed version of \gls*{dansep}, where all updating nodes in a full update cycle estimate the target signal of a fixed node instead of their own. We will show that this does not affect the \gls*{dansep} trajectory either, thereby proving convergence of \gls*{dansep}. In Step 3, we will finally prove optimality of \gls*{dansep} based on the observation that the second modified version of \gls*{dansep} corresponds to a full cycle of the block coordinate descent algorithm.

\textbf{Step 1:}
Consider a modified algorithm keeping all aspects of the proposed \gls*{dansep} algorithm intact except for the fact that, at iteration $i$ (and only at iteration $i$), the updating node $k$ aims at estimating the target signal at another node $\nu\in\K\backslash\{k\}$, i.e., $\dk[\nu]$ instead of $\dk$.
Instead of~\eqref{eq:lmmseDANSEplus}, the modified filter $\tWu$ of node $k$ is then obtained as:

\begin{equation}\label{eq:lmmseDANSEplus_alt}
    \begin{bmatrix}
        \vspace{.35em}
        \Wkku^{i+1}\\
        \Gkqu[k][l_1]^{i+1}\\
        \vdots\\
        \Gkqu[k][l_B]^{i+1}
    \end{bmatrix} \triangleq  \underset{\mathbf{W}\in\C[{\tMk^i}][Q]}{\mathrm{arg\,min}}\:
    \E[
        {\left\|
            \dk[\nu] - \mathbf{W}^\Her\begin{bmatrix}
                \yk\\
                \ettkq[l_1][k]^i\\
                \vdots\\
                \ettkq[l_B][k]^i
            \end{bmatrix}
        \right\|^2_2}
    ],
\end{equation}

\noindent
where $\{l_1,\dots,l_B\} = \Uk^i$ is the set of neighbors of node $k$ at iteration $i$. In this appendix, an underbar indicates that a vector or matrix is different in the modified algorithm compared to the original algorithm.
From the target signal definition~\eqref{eq:dk_to_sk}, again assuming $Q=S$ and $\Psibkov[q]$ full-rank and thus invertible$\fa q\in\K$, it holds that:

\vspace{-1em}

\begin{equation}
    \dk[\nu] = \Psibkov[\nu]\left(
        \Psibkov
    \right)^{-1}\dk \triangleq \Psibkq[k][\nu]\dk\fa (k,\nu)\in\K^2,\label{eq:dq_to_dk}
\end{equation}

\noindent
such that the solution of~\eqref{eq:lmmseDANSEplus_alt} is linked to that of~\eqref{eq:lmmseDANSEplus} as:

\vspace{-1em}

\begin{align}\label{eq:Wkk_alt}
    \Wkku^{i+1} &= \Wkk^{i+1}\Psibkq[k][\nu]^\Her\\
    \text{and}\:\:
    \Gkqu[k][l]^{i+1} &= \Gkq[k][l]^{i+1}\Psibkq[k][\nu]^\Her\fa l\in\Uk^i.\label{eq:Gltok_alt}
\end{align}

\noindent
Consequently, using~\eqref{eq:fusionrule} and~\eqref{eq:Gltok_alt}, the transformation matrix at non-updating nodes $q\in\K\backslash\{k\}$ is given as:

\vspace{-1em}

\begin{align}\label{eq:fusionrule_alt}
    \Tku[q]^{i+1} &\triangleq  \Tk[q]^{i}\Gkqu[k][n^i(q)]^{i+1}\\
    &= \Tk[q]^{i}\Gkq[k][n^i(q)]^{i+1}\Psibkq[k][\nu]^\Her\\
    &= \Tk[q]^{i+1}\Psibkq[k][\nu]^\Her\fa q\in\K\backslash\{k\}.\label{eq:fusionrule_alt_2}
\end{align}

\noindent
i.e., the transformation matrix in the modified algorithm is also post-multiplied by $\Psibkq[k][\nu]^\Her$.
Using~\eqref{eq:Wkk_alt} and~\eqref{eq:fusionrule_alt_2}, the fusion matrices defined in~\eqref{eq:PkDef} are then similarly changed in the modified algorithm:

\vspace{-1em}

\begin{align}
    \Pku[q]^{i+1}
    &= \begin{cases}
        \Pk[q]^{i}\Tk[q]^{i+1}\Psibkq[k][\nu]^\Her&\forall\: q\in\K\backslash\{k\}\\
        \Wkk[q]^{i+1}\Psibkq[k][\nu]^\Her&\text{for $q=k$}
    \end{cases}\label{eq:Pku_alt_1}\\
    &= \Pk[q]^{i+1}\Psibkq[k][\nu]^\Her\fa q\in\K.\label{eq:Pku_alt_2}
\end{align}

Now consider iteration $i+1$ of the modified algorithm where node $k'\neq k$ is the updating node (again, considering sequential node-updating). Because of~\eqref{eq:Pku_alt_2}, the partial in-network sum received by node $k'$ from its neighbor $l$, as defined in~\eqref{eq:fusionflow}, becomes:

\vspace{-1em}

\begin{align}
    \ettkqu[l][k']^{i+1}
    &=
    \Psibkq[k][\nu]\ettkq[l][k']^{i+1}\fa l\in\Uk[k']^{i+1}.\label{eq:fusionflow_alt_last}
\end{align}

\noindent
Then, at iteration $i+1$, node $k'$ estimates its own target signal $\dk[k']$ by computing the filter:

\begin{align}\label{eq:lmmseDANSEplus_alt_kprime}
    \begin{bmatrix}
            \vspace{.35em}
            \Wkku[k']^{i+2}\\
            \Gkqu[k'][l_1]^{i+2}\\
            \vdots\\
            \Gkqu[k'][l_B]^{i+2}
        \end{bmatrix}
    &= \underset{\mathbf{W}\in\C[{\tMk[k']^{i+1}}][Q]}{\mathrm{arg\,min}}\:
    \E[
        {\left\|
            \dk[k'] - \mathbf{W}^\Her\begin{bmatrix}
                \yk[k']\\
                \ettkqu[l_1][k']^{i+1}\\
                \vdots\\
                \ettkqu[l_B][k']^{i+1}\\
            \end{bmatrix}
        \right\|^2_2}
    ],
\end{align}

\noindent
where $\{l_1,\dots,\l_B\} = \Uk[k']^{i+1}$. 
By setting $\Gkqu[k'][l]^{i+2} = \Psibkq[k][\nu]^{-\Her}\Gkq[k'][l]^{i+2}\fa l\in\Uk[k']^{i+1}$, the effect of the pre-multiplication by $\Psibkq[k][\nu]$ in~\eqref{eq:fusionflow_alt_last} can be easily compensated for:

\begin{align}
    (\Gkqu[k'][l]^{i+2})^\Her\ettkqu[l][k']^{i+1} &= (\Gkq[k'][l]^{i+2})^\Her\Psibkq[k][\nu]^{-1}\Psibkq[k][\nu]\ettkq[l][k']^{i+1} = (\Gkq[k'][l]^{i+2})^\Her\ettkq[l][k']^{i+1}.\label{eq:Gkq_alt_2}
\end{align}

\noindent
Since the filter $\Wkk[k']^{i+2}$ applied to the local sensor signals $\yk[k']$ of node $k'$ at iteration $i+1$ is unaffected by~\eqref{eq:fusionflow_alt_last}, the filter $\tWu[k']^{i+2}$ is essentially unchanged in the modified algorithm. This also means that the transformation matrices at iteration $i+2$ are unchanged in the modified algorithm:

\vspace{-1em}

\begin{align}
    \Tku[q]^{i+2} &\triangleq  \Tku[q]^{i+1}\Gkqu[k'][n^{i+1}(q)]^{i+2}\\
    &= \Tk[q]^{i+1}\Psibkq[k][\nu]^\Her\Psibkq[k][\nu]^{-\Her}\Gkq[k'][n^{i+1}(q)]^{i+2}\\
    &= \Tk[q]^{i+2}\fa q\in\K\backslash\{k'\}.
\end{align}

\begin{figure}[h]
    \centering
    \includegraphics[width=\columnwidth,trim={0 0 0 0},clip=false]{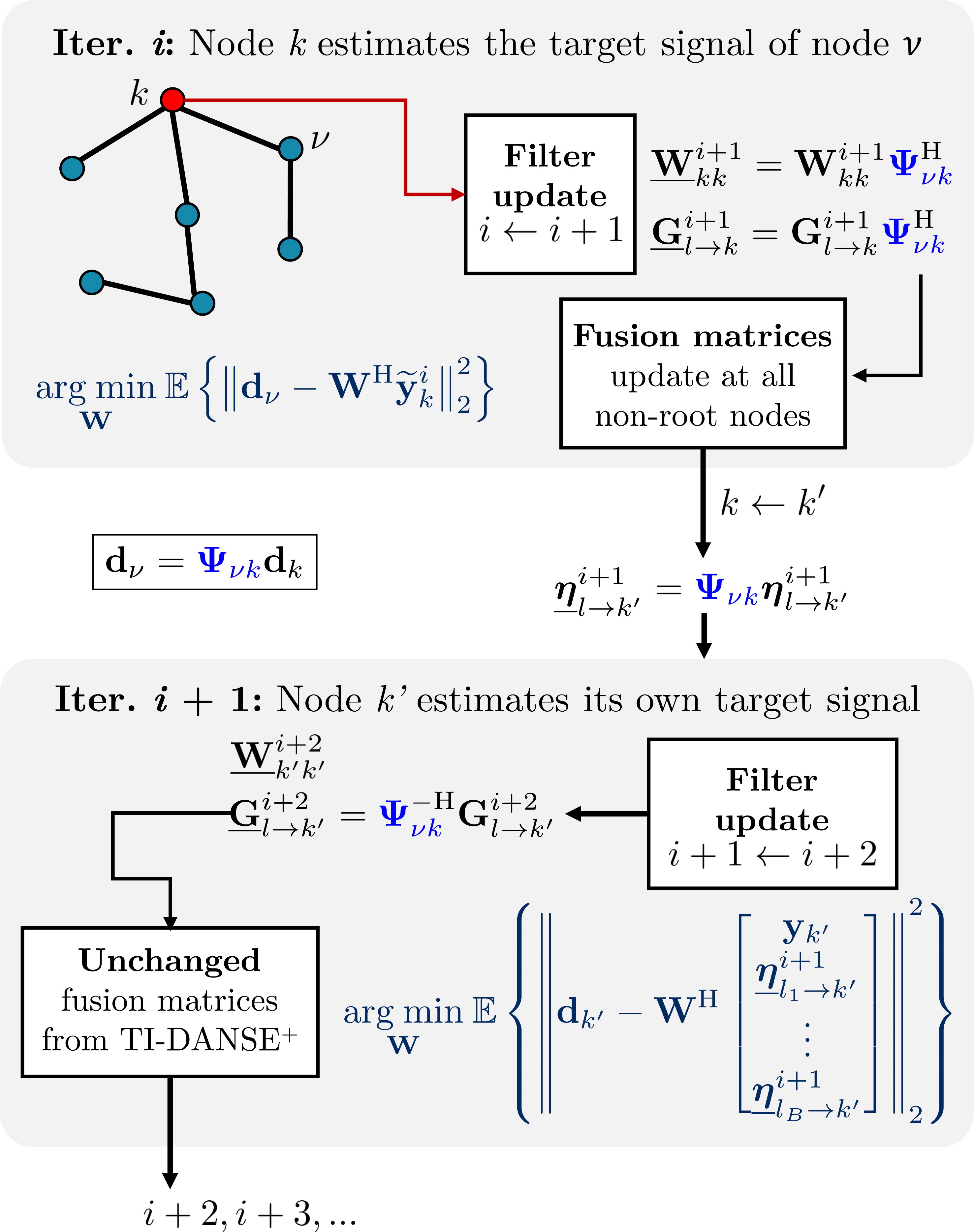}
    \caption{\change{
        Illustration of the the modified \gls*{dansep} algorithm used in Step 1 of the proof of Theorem 2. After iteration $i+1$, the impact of node $k$ estimating $\dk[\nu]$ instead of $\dk$ at iteration $i$ is compensated for, and the trajectory of \gls*{dansep} remains unaffected.
    }}
    \label{fig:proof_illus}
\end{figure}

\noindent
Therefore, from iteration $i+2$ onwards, the transformation matrices and hence the fusion matrices remain exactly the same in the modified algorithm as in the original \gls*{dansep} algorithm, as if node $k$ had estimated its own target signal $\dk$ instead of the target signal $\dk[\nu]$ of another node at iteration $i$. This shows that if node $k$ estimates $\dk[\nu]$ instead of $\dk$ in iteration $i$ and this is compensated for in the next iteration, the trajectory of \gls*{dansep} remains unaffected.
Note that $\Wkku^{i+2}=\Wkku^{i+1}$ continues to differ from $\Wkk^{i+2}$. However, this difference disappears once node $k$ becomes the updating node again, i.e., after a full update cycle.
\change{\figref{fig:proof_illus} illustrates the above reasoning.}

Going back to the original, unmodified \gls*{dansep} algorithm, suppose that node $\nu$ is the updating node at iteration $i+1$. The previous reasoning implies that the result of this \gls*{dansep} update is the same as if the previous updating node (at iteration $i$) also had updated its filters towards estimating $\dk[\nu]$.
Considering the \gls*{mse} cost function at node $q$, i.e., $J_q(\mathbf{W})\triangleq \El[{
    \dk[q] - \mathbf{W}^\Her\mathbf{y}
}]\fa q\in\K$, we have that, with $\nu$ as the updating node at iteration $i+1$:

\begin{equation}
    J_\nu\left(
        \Wk[\nu]^{i+1}
    \right)\leq J_\nu\left(
        \Wk[\nu]^{i}
    \right),
\end{equation}

\noindent
where $\Wk[\nu]^i$ is the network-wide \gls*{dansep} filter at iteration $i$ at node $\nu$ as defined in~\eqref{eq:nw_filter_dansep}.

\textbf{Step 2:}
Consider now a second modified version of \gls*{danse} where all updating nodes $k\in\K$ in a full update cycle estimate $\dk[\nu]$ (where $\nu$ is a fixed node) instead of $\dk$. Letting $j$ and $j+K$ denote iterations where $\nu$ is the updating node, the reasoning from step 1 can be applied to all iterations between $j$ until $j+K$, i.e., for a full update cycle. The trajectory of the \gls*{dansep} algorithm consequently remains unaffected also with this further modification.

Once again going back to the unmodified \gls*{dansep} algorithm, every time node $\nu$ updates its filters, the result is thus as if all previously updating nodes in the update cycle had also updated their filters towards estimating $\dk[\nu]$. This leads to:

\vspace{-1em}

\begin{equation}
    J_\nu\left(
        \Wk[\nu]^{j+K}
    \right)\leq J_\nu\left(
        \Wk[\nu]^{j+K-1}
    \right)\leq \dots \leq J_\nu\left(
        \Wk[\nu]^{j}
    \right).
\end{equation}

\noindent
The same reasoning can be applied to any node $q\in\K$ and any corresponding iteration $j\in\N$ where $q$ is the updating node. Therefore, all sequences $(J_q(\Wk[q]^{j}))_{j\in\N}\fa q\in\K$ are decreasing sequences. Since they have a lower bound, they converge to a limit. This proves convergence of the \gls*{dansep} algorithm.

\textbf{Step 3:}
A full update cycle of the modified \gls*{dansep} algorithm where all updating nodes estimate $\dk[\nu]$ (with $\nu$ a fixed node) corresponds to one full cycle of the block-coordinate descent algorithm, i.e., to a nonlinear Gauss-Seidel~\cite{bertsekasParallelDistributedComputation2015} algorithm for a fixed \gls*{lmmse} problem estimating $\dk[\nu]$, where alternating blocks of variables are optimized (at updating node $k$, where index $k$ changes at every iteration in a round-robin fashion) while the other variables (at non-updating nodes) are fixed. Such algorithm is known to converge Q-linearly to the optimal solution. Since the trajectory of the \gls*{dansep} algorithm remains unaffected by the modification made above, the \gls*{dansep} algorithm converges Q-linearly to the optimal solution of the \gls*{lmmse} problem at each node.
\hfill$\blacksquare$

\AtNextBibliography{\footnotesize}
{
\printbibliography
}

\end{document}